\documentclass{emulateapj}
\slugcomment{To be submitted to the Astrophysical Journal}
\shorttitle{The Coronae of AB Dor and V471 Tau}
\shortauthors{Garc\'{\i}a-Alvarez et al.}
\begin{document}
\title{The Coronae of AB Dor and V471 Tau: Primordial
angular momentum vs tidal spin-up}
\author{David Garc\'{\i}a-Alvarez, Jeremy J. Drake, LiWei Lin, Vinay L. Kashyap, B. Ball }
\affil{$^1$Harvard-Smithsonian CfA, \\ 60 Garden Street, \\ Cambridge, MA 02138}

\begin{abstract}
The zero-age main-sequence star AB~Dor and the K dwarf component of
the V471~Tau close binary have essentially identical rotation rates and
spectral types.  An analysis of their high resolution {\it Chandra}
X-ray spectra reveals remarkably similar coronal characteristics in
terms of both temperature structure and element abundances.  Both
stars show depletions of low FIP elements by factors of $\sim 3$, with
higher FIP elements showing more mild depletions.  No evidence for
enhancements of very low FIP ($< 7$~eV) elements, such as Na, Al and
Ca, as compared to other low FIP elements was found.  The abundance
anomaly pattern for AB~Dor and V471~Tau is similar to, though less
extreme than, the abundance anomalies exhibited by active RS-CVn-type
binaries.  While we find statistically significant structure in the
underlying differential emission measure distributions of these stars
over narrow temperature intervals, this structure is strongly dependent
on the lines used in the analysis and is probably
spurious.  Based on their X-ray similarities, we conclude that the
exact evolutionary state of a star has little effect on coronal
characteristics, and that the parameters that dominate coronal
structure and composition are simply the rotation rate and spectral
type.
\end{abstract}


\section{Introduction}

%Prior to the {\it{Chandra}} and XMM-{\it{Newton}} era, with the
%exception of a single, low signal-to-noise, {\it Einstein} grating
%observation of Capella \citep{Vedder??}, 
%coronal observations capable of resolving
%individual spectral lines in the X-ray range had been performed only
%for the Sun.  

The solar coronal abundance anomaly commonly known as the ``FIP
Effect'' (or First Ionization Potential Effect), in which low first
ionization potential (FIP) elements ($\leq$\,10\,eV; e.g. Si, Fe, Mg)
appear enhanced by average factors of 3-4, was already well
established by the time the first observational clues to similar
abundance anomalies in stellar coronae were uncovered in the 1990's
\citep[e.g.][]{Feldman92}.  These clues came from low resolution soft
X-ray spectra ({\it GINGA, ASCA, BeppoSAX}), together with moderate
resolution extreme ultraviolet (EUV) spectra obtained by the Extreme
Ultraviolet Explorer (EUVE).  The early stellar studies found evidence
for a solar-like FIP Effect in some stars, but for the more active
stars the abundances pointed toward metal-paucity rather than
enhancement \citep[e.g.][ and earlier references therein]{SDrake96}.  

In the last three years---the beginning
of the {\it Chandra} and {\it XMM-Newton}---era has seen the early
hints of abundance anomalies fleshed out into an interesting array of
diverse abundance patterns in which active stars appear to show signs
not only of low FIP element depletion, but also of high FIP element
enhancements \citep[][]{Singh95}.
Studies of solar-like active stars confirm the suspicions engendered
by earlier work \citep[e.g.][ and earlier references therein]{Drake96} that a transition from
a metal-depleted to a metal-rich corona occurs as the activity decreases
\citep{Guedel02}; this is now better characterised as a change from an 
apparently ``inverse FIP Effect'' to a FIP Effect. 
\citet{Audard03} showed a similar transitions from an 
inverse FIP effect to the absence of an obvious FIP effect for RS CVn binary
stars with decreasing activity.  Only one star studied in detail to date based on high
resolution spectra appears to have a coronal composition demonstrably similar
to that of its underlying photosphere---the subgiant Procyon (F5IV) 
\citep{Drake95,Raassen02}.

%element abundance anomalies in coronal
%plasma such as metal paucity \citep[e.g.][and references
%therein]{Drake01}. However, the reliability of the element abundances
%thus obtained is questionable due to the global model fitting methods
%that must be used to model these spectra
%\citep[e.g.][]{Jordan98,Drake01,Drake02}. High resolution X-ray
%spectra can currently provide detailed and reliable abundance for
%individual elements, while simultaneously giving us new insights and
%diagnostics on coronal structure and dynamics.

In the last decade, the community working on coronal abundances 
have argued \citep[e.g.][ and earlier references
therein]{Drake03a} that coronal abundances, when better understood,
might provide new and powerful diagnostics of the physical processes
underpinning stellar coronae.  The emerging patterns of coronal
abundance anomalies are telling us something about the dynamics
structure and heating of coronal plasma; the challenge is in learning
to read these patterns.

Aiming toward this goal, one question that
arises is that of fine tuning of abundance patterns: how similar are
abundance anomalies in stars that superficially might have similar
parameters?  \citet{Drake03b} presented a differential analysis of the
active binaries Algol and HR1099---two binary systems whose
coronally-dominant components are of very similar spectral type and
rotation period, but which have had quite different evolutionary
histories---and found strong similarities in their quiescent X-ray
spectra and coronal abundances.  

Here, we note that the young single active star AB Dor and the active
component of the binary system V471 Tau are both early K dwarfs and
have rotation periods of half a day.  Superficially, we might also
expect them to have very similar coronal characteristics.  However,
\citet{Still03} have derived coronal abundances for V471 Tau based on
low resolution ASCA CCD pulse height spectra which are significantly
larger than those derived for AB Dor by other authors using similar
observational material, in addition to EUVE, XMM-Newton and {\it
Chandra} spectra
\citep[][]{Rucinski95,Mewe96,Ortolani98,Maggio00,Guedel01,Sanz-Forcada03}.
The apparent difference between the coronal abundances of the two
stars might be hinting that their coronae are indeed fundamentally
different.  Such a dissimilarity might reflect the entirely different
angular momentum evolution of the two stars: the rapid rotation of
AB~Dor is due to its relative youth, while that of V471 Tau is due to
tidal spin-up from interaction with its close companion.

In this paper, using {\it Chandra} High and Low Energy Transmission
Grating spectrograph (HETGS and LETGS) observations, we present a
comparative analysis of the coronal X-ray spectra of AB Dor and V471
Tau, with particular emphasis on their abundances.  We first describe
the two stars and briefly review earlier work (\S\ref{s:stars}), then
in \S\ref{s:obs} we report on the {\it Chandra} observations and data
reduction.  The methods used for a differential emission measure
analysis together with results obtained are shown in
\S\ref{s:anal}. In \S\ref{s:results} we discuss our results on the
coronal and temperature structure and report our conclusions in
\S\ref{s:concl}.

\section{Program Stars}
\label{s:stars}
\subsection{AB Dor}
\label{s:abdor}
AB Doradus (HD 36705) is a young, relatively bright (V=6.9) and
rapidly rotating late-type star with spectral type K0-2\,V. It is an
example of the very active cool stars that are just evolving onto the
main sequence. Its age is estimated to be 20-30 Myr
\citep{Guirado97,Collier-Cameron97}, and \citet{Innis86} suggested 
that AB Dor is member of the Pleiades based on its space velocity. 
The main physical parameters of
AB Dor are shown in Table~\ref{t:par}. \citet{Amado01} showed evidence
of a solar-type activity cycle on AB Dor, with the starspot coverage
increasing and decreasing with a period of 22-23 yr. Extensive Doppler
imaging on this star has revealed the presence of a long term polar
spot \citep{Collier-Cameron94,Donati97b,Donati99b}. The very fast
rotation rate of AB Dor produces strong magnetic fields which has been
investigated through Zeeman-Doppler Imaging by several authors
\citep[e.g.][]{Donati97b,Donati99b}. \citet{Lim94} presented measurements of the spectrum and polarization of the rotationally modulated radio emission on AB~Dor. 

AB Dor was originally discovered as a bright X-ray source, with {\it
Einstein}, by \citet{Pakull81}. Since then it has been extensively
observed in the ultraviolet \citep{Rucinski95,Moos00,Vilhu01} and
X-rays
\citep{Vilhu87,Collier-Cameron88,Mewe96,Kuerster97,Maggio00,Guedel01}.
\citet{Kuerster97} found variability on short timescales but no
long-term trend or cyclical variability of the X-ray emission. Their
study showed that about 15\%\ of the X-ray emission is rotationally
modulated, indicating that at least part of the X-ray emission must
have scale sizes below a stellar radius. \citet{Collier-Cameron88}, \citet{Schmitt98} and \citet{Maggio00} have reported X-ray flares on AB Dor. 

The photospheric abundances of AB Dor are relatively well studied; 
\citet[][]{Vilhu87} found a solar-like metallicity,
[Fe/H]=0.1$\pm$0.2, based on a high resolution spectroscopic study. The Fe photospheric abundance reported for the Pleiades by several authors \citep[e.g.;][]{King00,Boesgaard89b,Cayrel85} show a slightly super-solar value, 0.02$<$[Fe/H]$<$0.15. Similar values have been reported for other elements ([O/H]=0.14 \citep{Schuler04}, [Mg/H]=0.05 and [Si/H]=0.07 \citep{King00}). A low coronal iron abundance, [Fe/H]=-0.95, found by
\citet{Mewe96} using simultaneous observations with EUVE and ASCA, was
confirmed by \citet{Guedel01} using {\it{XMM-Newton}} data and what
appears to be an ``inverse FIP'' effect---that elements with high
First Ionization Potentials (FIPs) (e.g., C, N, O, Ne, Ar) appear
enhanced relative to low FIP elements (e.g., Mg, al, Si, Fe), opposite
what is observed in the solar corona \citep[e.g.][]{Feldman92}. These
values were 3-5 times lower than the photospheric abundance obtained
by \citet{Vilhu87}. \citet{Sanz-Forcada03}
reported coronal abundances for AB~Dor based on XMM-Newton and {\it Chandra}
spectra showing a similar Fe depletion.

\subsection{V471 Tau}
\label{s:v471tau}
V471 Tau (BD+16 516) is a rapidly rotating relatively faint (V=9.5)
detached eclipsing binary first observed by \citet{Nelson70}. The
system is composed of a cool main-sequence chromospherically active
K2\,V star \citep{Guinan84} and a degenerate hot white dwarf (DA1.5)
separated by only 3.4 stellar radii. During the primary eclipse, the
hot white dwarf acts as a beaming probe for the K2 star's atmosphere
\citep{Guinan86}. The system is thought to belong to our nearest open
cluster the Hyades based on Hipparcos proper-motion and parallax
measurements \citep{Provencal96,deBruijne01}, although this point has
been controversial \citep[see discussion by][]{Martin97}. The basic
properties of V471 Tau are summarized in Table~\ref{t:par}.

As discussed by \citet{Paczynski76}, the V471 Tau system has probably
been through a common-envelope (CE) stage, and the system is evolving
to become a cataclysmic variable in which the K dwarf overfills its
Roche lobe, leading to mass transfer onto the white dwarf. V471 Tau is
often considered to be the prototypical "precataclysmic" binary
\citep[see][]{Bond85}; it is one of two known post-CE, pre-CV systems
in the Hyades cluster (the other being HZ 9). Based on the same {\it
{Chandra}} LETGS spectrum that we analyse in this paper,
\citet{Drake03c} provided the first direct observational evidence of
the CE phase of 
V471 Tau, demonstrating that the coronal C abundance was depleted
relative to N as would be expected from the surface contamination of
the K dwarf by the evolved and expanded envelope of the primary star. 
V471~Tau is currently the
only known precataclysmic variable detected at radio
wavelengths \citep{Lim96}.

Several observers have reported a strong, complex, and variable
H$\alpha$ feature that goes from absorption to emission as the system
rotates \citep{Young88,Skillman88}. Other chromospheric lines such as the \ion{Ca}{2} H \& K and \ion{Ca}{2} infrared triplet show similar
behavior confirming that the activity is coming from the chromosphere
of the K dwarf. Different authors have presented evidence for
magnetically induced activity on the K dwarf based on optical and
photometric flares \citep{Young91}, X-ray flares \citep{Young83} and
radio flaring activity \citep{Patterson93,Nicholls99}.
\citet{Jessen86} and \citet{Sion98} found a periodic 9.25 minute
modulation in the soft X-ray and EUV bands of V471 Tau, subsequently
found in the optical band by \citet{Robinson88}. It is now believed
that the variability is caused by rotational modulation of the
magnetic white dwarf \citep{Clemens92,Barstow93}.  \citet{Wheatley98}
found that the rotation rate and X-ray luminosity of the K star in
V471 Tau are typical of K stars found in the Pleiades.

The photospheric abundances for the Hyades open cluster, to which V471
Tau is thought to belong, are generally better determined than those
for most coronally active field stars.
\citet{Cayrel85} reported an analysis of high
signal-to-noise ratio Coud\'e spectra for 12 Hyades dwarfs, together
with solar spectra for comparison, in order to obtain accurate
abundances for iron and other metals in the Hyades. They determined 
mean Fe and Si abundances slightly higher
than the solar values, [Fe/H]=0.12$\pm$0.03 and [Si/H]=+0.16$\pm$0.04.
\citet{Varenne99} derived abundances for
\ion{O}{1}, [O/H]=+0.07 for a similar sample. \citet{Boesgaard89} and
\citet{Varenne99} found, based on a sample of F stars in the Hyades,
similar values for the [Fe/H]. \citet{Smith97} reported a global metallicity 
for the Hyades cluster of [Fe/H]=+0.13 based on values determined for
individual dwarfs and giants from high resolution spectra. \citet{Smith99} 
derived similar values for the Hyades K0 giants $\gamma$ Tau and $\epsilon$ Tau.
Based on low resolution X-ray CCD spectra, 
\citet{Still03} reported coronal abundances 
showing only a very mild Fe abundance depletion relative to
photospheric estimates, together with evidence for an inverse FIP
effect. In their analysis Fe coronal abundance shows a factor of $\sim$2 depletion while Ne shows a factor of $\sim$3 enhancement, both relative to photospheric values.

\section{Observations}
\label{s:obs}

The {\it{Chandra}} HETGS observation of AB Dor was carried out in 1999
September using the AXAF CCD Imaging Spectrometer (ACIS-S), while the
{\it{Chandra}} LETGS observation of V471 Tau was carried out in 2002
January using the High Resolution Camera spectroscopic (HRC-S)
detector.  Both observations employed their respective detectors in
their standard instrument configurations. The observations are
summarized in Table~\ref{t:obs}.

Pipeline-processed (CXC software version 6.3.1) photon event lists
were reduced using the CIAO software package version 2.2, and were
analyzed using the IDL\footnote{Interactive Data Language, Research
Systems Inc.}-based PINTofALE software suite \citep{Kashyap00}.  In
the case of V471~Tau, processing included filtering of events based on
observed signal pulse-heights to reduce background (Kashyap \& Drake
2000, Wargelin et al., in preparation).  The analysis we have
performed consisted of line identification and fitting, reconstruction
of the plasma emission measure distribution including allowance for
blending of the diagnostic lines used, and finally, determination of
the element abundances.

The white dwarf component of V471 Tau exhibits a strong thermal
continuum at long wavelengths ($>$ 50\AA), which is clearly seen in
the LETG spectrum.  This was studied in previous EUV observations
\citep[e.g.][]{Barstow92}.  We note, however, that this continuum
contributes no significant flux shortward of 50\AA.

\begin{figure}
%\begin{turn}{90}
\plotone{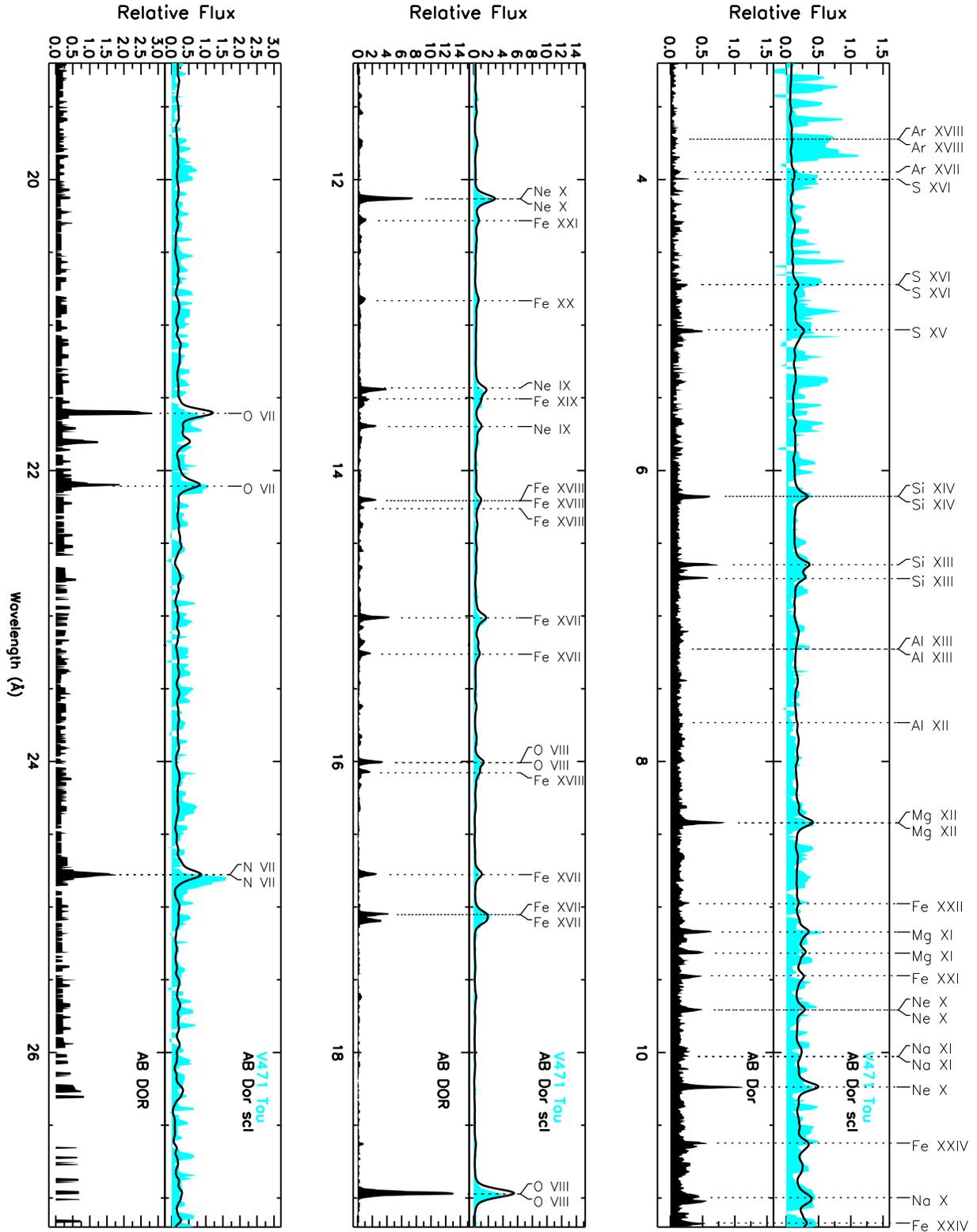}
%\end{turn}
\caption{{\it Chandra} X-ray spectra of AB Dor (black shade lower panel; HETG+ACIS-S)
and V471 Tau (grey shade upper panel; LETG+HRC-S) respectively. The strongest lines over the observed wavelength range are identified. The thick black line in the upper panels shows AB Dor spectra after smoothing with a beta-profile in order to match with the resolution of the V471 Tau spectrum. Note the strong correspondence between the two spectra. 
}
\label{f:sp}
\end{figure}

Fig.~\ref{f:sp} shows the {\it Chandra} X-ray spectra of AB Dor (black
shade; HETG+ACIS-S) and V471 Tau (grey shade; LETG+HRC-S) in the limited
wavelength range 3-27~\AA\ which contains the prominent lines of Na,
N, O, Ne, Mg, Fe, S and Si. 
The strongest coronal lines are identified.  The thick
black line shows the AB Dor spectrum after smoothing in
order to match the resolution of the V471 Tau spectrum.  
The spectra of both stars show a
remarkable similarity of lines, both in terms of which lines are prominent,
from H- and He-like ions and the broad range of charge states of Fe,
and in their intensities.  Figure~\ref{f:sp} shows that at shorter
wavelengths the spectrum is dominated by emission lines that we
interpret as coronal emission from the K dwarf component.

According to the ephemeris of \citet{Guinan01}, the LETGS observation of V471 Tau started at phase $\phi=0.05$ and ended at $\phi=1.97$, where
$\phi=0$ when the K dwarf component is nearest the observer.

\section{Analysis}
\label{s:anal}
\subsection{Photometry}

\begin{figure}
%\begin{turn}{90}
%\plotone{/export/dgarcia/data/abdor/primary/ABDOR_V471tau_wide.ps}
\plotone{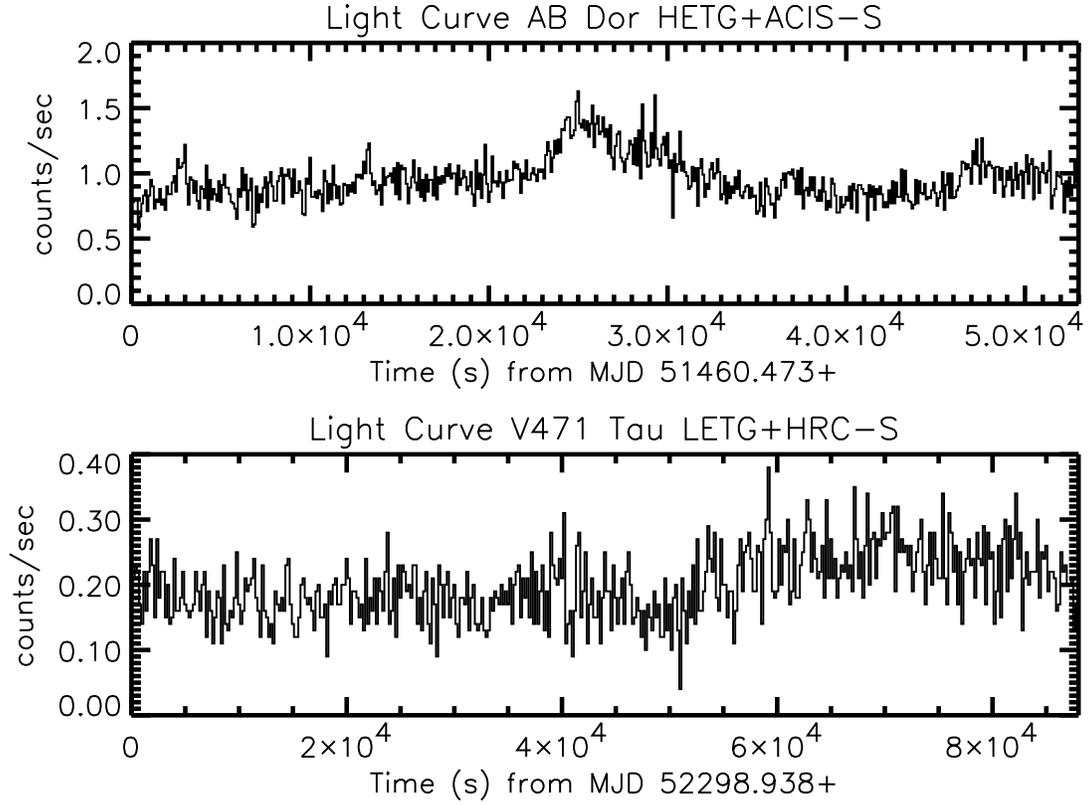}
%\end{turn}
\caption{{\it Chandra} X-ray light curves of AB Dor (top; HETG+ACIS-S)
and V471 Tau (bottom; LETG+HRC-S) respectively. The AB Dor and V471 Tau light curves are binned at 100s and 200s intervals respectively. Both objects were relatively
quiescent, showing no large flare events. Based on the ephemeris of
\citet{Guinan01}, the V471 Tau light curve spans
phases $\phi=0.05$ to 1.97.}
\label{f:lc}
\end{figure}

Before commencing spectral analysis, we first checked for flare
activity that could affect not only the shape of the differential
emission measures (DEMs) but might also be accompanied by detectable
changes in the coronal abundances of the plasma that dominates
disk-averaged spectra \citep[see,e.g.][]{Ottmann96,Mewe97,Ortolani98,Favata99,Favata00,Maggio00,Guedel01}.
Light curves for the AB Dor and V471 Tau observations were derived
using the 0th order events extracted in a circular aperture, in the
case of V471~Tau, and using the dispersed photon events and excluding
0th order (to avoid photon pile-up) in the case of AB~Dor.  Events were
then binned at 100s and 200s intervals for AB Dor and V471 Tau,
respectively.  Fig.~\ref{f:lc} shows the light curves for both
observations; note that both light curves are relatively flat and
devoid of large flare activity.

Several authors have reported coronal rotational modulation on active
stars \citep[e.g.][]{Drake94,Audard01,Garcia-Alvarez02b}, including
AB~Dor \citep{Kuerster97}.  However, as is shown in Fig.~\ref{f:lc}
(bottom panel), no obvious modulation of the coronal emission with
orbital phase was seen on either star, though in the case of AB~Dor,
the observation only covers slightly more than one rotational period.
In the case of V471~Tau, modulation due to modification of the
underlying chromospheric structure might be expected, for example, if
the part of the K dwarf corona irradiated by the white dwarf were affected by
the UV radiation field of the latter.  This is apparently not
the case to any significant extent. A similar conclusion has also been
reached by \citet{Wheatley98}.

%In the case of
%AB~Dor, convincing rotational modulation of coronal emission has also
%proved elusive, even during a long EUVE observation spanning ??ks
%\citep{Steve White conf paper??}.

We conclude that both AB~Dor and V471~Tau observations are
representative of the stars during times of quiescence, and therefore
treat the observations in their entirety for the remainder of the
analysis.  

%It is also worth noting the orbital phase of V471 Tau
%during the observations: according to the ephemeris of
%\citet{Guinan01}, the observation started at phase $\phi=0.05$ and
%ended at $\phi=1.97$, where $\phi=0$ when the K dwarf component is
%nearest the observer.

\subsection{Spectroscopy}

Spectral line fluxes for both AB Dor and V471 Tau were measured by
fitting modified Lorentzian (``$\beta$-profile'') functions of the form
\begin{equation}
F(\lambda)=a/(1+(\frac{\lambda-\lambda_0}{\Gamma})^2)^\beta
\end{equation}
where $a$ is the amplitude and $\Gamma$ a characteristic line width.
For a value of $\beta=2.4$, it has been found that this function
represents the {\it{Chandra}}
transmission grating instrumental profile to within photon counting
statistics for lines with a few
1000 counts or less \citep{Drake03d}.   Spectra line fitting was undertaken using
the FITLINES utility in the PINTofALE spectral analysis package 
\citep{Kashyap00}.\footnote{PINTofALE is freely available from 
http://hea-www.harvard.edu/PINTofALE/}
In the
case of obvious blends, as in the \ion{Ne}{9}-\ion{Fe}{19} 13~\AA\ region, 
we have performed multi-component fitting.

While strong lines can be easily identified, the sum
of weak lines, each of which can be undetectable, can produce a
``pseudo-continuum''. The true continuum level was set using the
spectral regions 2.4-3.4, 5.3-6.3 and 19.0-20.0~\AA, which we judged
to be essentially ``line-free'', based on both visual inspection and
the examination of radiative loss models.  These continuum points were
then used to normalize model continua computed using a test
Differential Emission Measure taken from an analysis of a typical 
active star (AU~Mic in this case, \citet{Drake04}).  In
principle, the fluxes should be re-measured once the final DEM has
been determined; in practise, there was little difference in
the {\em shapes} of trial and final continuum models so that this
additional iterative step was not required.

\begin{figure}
%\begin{turn}{90}
\plotone{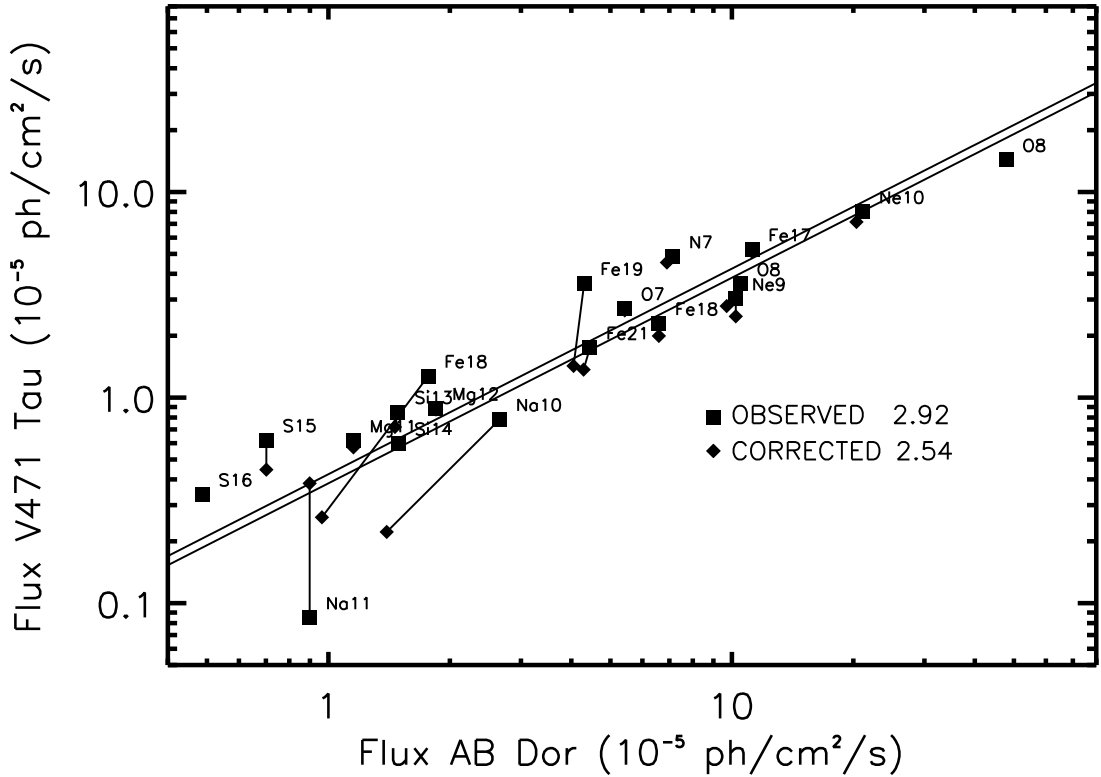}
%\end{turn}
\caption{Comparison of observed and blend corrected fluxes of AB Dor and V471 Tau {\it Chandra} X-ray spectra. Note that the fluxes are corrected for distance.}
\label{f:flx}
\end{figure}

The emission measure distribution and abundance analysis employed the
CHIANTI database \citep{Dere01} and the ionization balance of
\citet{Mazzota98}, as implemented in the PINTofALE software package
\citep{Kashyap00}. Table~\ref{t:flx} shows the measured fluxes of the
emission lines identified and used in this analysis. We have computed
correction factors (CF) for line blends by computing complimentary
error functions for each contaminant (i.e. Gaussian profiles are
assumed). A set of interesting lines and a set of contaminant lines, which 
contain necessary line information (emissivity, wvl, z, etc.), are used. The procedure identifies which lines are possible contaminants
for each analysed line. The contaminants must lie within two sigma on either side of the line and be greater than a user-defined threshold. The CF are defined as follows:
\begin{equation}
CF = {I(\lambda_i) \over I(\lambda_i) + \sum_{j \neq i} I(\lambda_j)}
\end{equation}
where I($\lambda_i$) = intensity of the line being contaminated by
lines $j \neq i$. CF
for line blends for the studied lines range from 0.89 to
0.99. Fig.~\ref{f:flx} shows the comparison of our observed and blend
corrected fluxes for both AB Dor and V471 Tau.  
Note that the fluxes are corrected for distance and exposure
time. 

\subsection{Differential Emission Measure}
\label{s:dems}
%\subsubsection{Introduction}

We adopted the following formalism for the Differential Emission
Measure.  The line flux, for
optically-thin plasma, for the transition of an ion between levels $i
\rightarrow j$ can be written as:
\begin{equation}
F_{ij} = AK_{ij} \int_{\Delta T_{ij}} G_{ij}(n_e,T) 
\Phi(T) \;dT
%\overline{N_e^2}(T)\, \frac{dV(T)}{dT} \;dT
\,\,\, \mbox{erg cm$^{-2}$ s$^{-1}$}
\label{e:flux}
\end{equation}
where A is the abundance, $K_{ji}$ is a known constant which includes the frequency of the
transition and the stellar distance, n$_e$ is the electron density,
G$_{ij}$ is the contribution function and T is the electron
temperature. The kernel $\Phi(T)$ is commonly known as the {\em
differential emission measure} (DEM) and is a measure of the amount of
emitting power---correlated with the amount of emitting material---as
a function of temperature in the coronal plasma. It is formally
defined as: 
\begin{equation}
\Phi(T)={n_e^2}(T)\frac{dV(T)}{dT}
\label{e:dem}
\end{equation} 
Two different approaches are commonly employed in the derivation of
the DEM in the corona, namely, line-based methods and global
fitting techniques.  For the latter the parameters are determined by a
$\chi^2$ minimization of the observed and trial model spectra.  We
prefer the former method for the greater control over the
spectroscopic diagnostics entering the analysis that it allows.

A given line flux depends on both temperature and
on the abundance of the element in question.  For this reason, methods
wishing to take advantage of bright lines from a number of different
elements generally have to perform simultaneous modelling of the DEM
and of abundances \citep[see, e.g.][]{Maggio04,Argiroffi04}.  
However, the ratio of two emission lines from ions of the
{\em same} element is independent of the abundance of the chosen
element.  We have therefore devised a method that uses line {\em
ratios} instead of line fluxes directly.  A similar method 
was developed independently by \cite{Schmitt04} who also discuss the 
benefits of such an approach.  Our basic set of diagnostics
comprises the H-like/He-like
resonance line flux ratios for the elements O, Ne, Mg, and Si,
line ratios involving Fe~XVII, Fe XVIII and Fe~XXI
resonance lines, and measurements of the continuum flux at points in
the spectrum that are essentially free of lines (the same regions
noted above with regard to spectral line intensity measurement). Our
method is described in more detail by \citep{Drake04}.

%Several
%authors have previously used this line ratio concept to determine DEM
%in both solar and stellar coronae
%\citep[e.g.][]{McKenzie92,Fludra95,McIntosh00,Schmitt02}. {\it{Chandra}}
%and XMM-{\it{Newton}} observations cover the stronger resonance lines
%of the hydrogen- and helium-like ions. 

The motivation underlying the development of this method lies
in a desire to survey coronal abundances and trends in temperature
structure independent from issues regarding atomic data.
\citet{Kashyap98} have demonstrated that a reconstructed DEM can be
quite sensitive to the lines used for the reconstruction, and also on
the atomic data adopted for those lines.  Our basic
line set comprises the brightest lines in stellar coronal spectra and
are easily measured in essentially all well-exposed {\em Chandra}
grating observations of stellar coronae, such that star-to-star
variations in diagnostic lines used can be avoided.  The H-like
and He-like ions are also likely to be represented by more accurate
theoretical line intensities than lines from more complex ions.
Similarly, the theoretical description of free-free and bound-free
continua used in our DEM analysis 
is thought to be well-understood.  In particular, we note
that the continua in these hot coronal sources tend to be dominated
by free-free emission.

\subsection{Algorithm and Line Selection}

In order to obtain the {\em differential emission measure}, $\Phi(T)$,
we have
performed Markov-Chain Monte-Carlo analysis using a Metropolis algorithm
(MCMC[M]) on the set of supplied line flux ratios
\citep{Kashyap98}.  Advantages of this method include the ability to
estimate uncertainties on the derived function $\Phi(T)$, and the
avoidance of unnecessary smoothing constraints.  
Eq.~\ref{e:dem} defines the DEM as a continuous
function of temperature $T$. 
The MCMC[M] method yields the emission
measure distribution over a pre-selected temperature grid, where the
DEM is described by a simple histogram. 
In our case, a set of temperatures $T_n$, with
$\Delta\,\log\,T[K]$=0.1 and ranging from $\log\,T$=6.1 to
$\log\,T[K]$=8.0, define the $\Phi(T_n)$. The derived $\Phi(T_n)$ is
only reliable over a certain temperature range if we have enough lines
with contribution function G($T_{max}$)$\sim$
G($T_{n}$).

The inversion problem is ill-conditioned which cannot be overcome by any fit procedure criteria applied \citep{Craig76}. Nevertheless, in order to ``avoid'' the problem of mathematical ill-conditioning, DEM
reconstruction techniques often rely on modeling the DEM as smooth
splines or polynomials, and are thus artificially constrained to be
smooth.  We do not impose arbitrary smoothing restrictions, but
instead let the algorithm find the best possible model allowed by the
data \citep[e.g.;][]{Kashyap98}, while applying a variable {\em local} 
smoothing to the DEM.  This local smoothing is based on the diagnostic
information available: if there are many spectral lines in a given
temperature range with different temperature sensitivities, smoothing
is essentially switched off; instead, more smoothing is applied where
diagnostic information is more sparse.  This technique 
often allows solutions containing structure over
small temperature ranges, whose formal significance may be judged by the
sizes of the error bars that the algorithm also produces.  While the
reality of strongly peaked features in stellar DEMs is still
in question, other techniques have found evidence for sharply peaked
structures in some coronae, as illustrated, for example, 
by the emission measure derived for Capella using EUVE data
\citep{Dupree93} and for the Sun using Yohkoh data \citep{Peres01}. Although we have found a ``good'' solution for the DEMs, by using the local smoothing criteria, one has to keep in mind that we cannot a priori exclude other solutions
in the parameter space that describe the spectra equally well.

Table~\ref{t:flx} shows the lines used
for the DEM(T$_n$) reconstruction.  In the X-ray range analyzed, the
lines with the coolest peak formation temperature are the resonance
lines \ion{O}{7} and \ion{N}{7}, $\log\,T[K]$=6.3, while the hottest
peak formation temperature is the resonance line \ion{S}{16},
$\log\,T[K]$=7.4. Based on the lines we use in our analysis we are
able to obtained a well-constrained DEM(T$_n$) between $\log\,T[K]$=6.3
and $\log\,T[K]$=7.4; larger uncertainties are obtained outside that
range.

In order to probe the consistency of the results, depending on the
line ratios used to derive the DEMs, we chose four different sets of
H-like, He-like and highly ionized Fe lines. Set 1 contains H-like and
He-like lines only (O, Mg, Ne, Si), Set 2 contains O, Mg, Ne, Si and
S, Set 3 contains O, Si, S, \ion{Fe}{17}, \ion{Fe}{18} and
\ion{Fe}{21} and Set 4 contains O, Ne, Mg, Si, \ion{Fe}{17},
\ion{Fe}{18} and \ion{Fe}{21}. Set 1 represents the basic core of
H-like and He-like lines used by \citet{Drake04} to derive DEMs. Set 2
adds sulfur to the previous set of lines, which allows us to increase
the upper temperature range up to $\log\,T[K]$=7.4, although the
significance and signal-to-noise for these lines in our spectra are low. 
Set 3 contains H-like and
He-like lines that constrain the structure of the DEM in the upper and
lower temperature range, while the iron ions give information about
the structure of the DEM in the intermediate temperature range. Set 4
contains all the ions analyzed. All the sets contain
\ion{O}{8}, which constrains the DEM in the lower temperature range,
$\log\,T[K]$=6.3. Note that the electron density (n$_e$) adopted for Eq.~\ref{e:dem} will not produce 
significant differences in the derived DEMs since the chosen lines are not n$_e$-sensitive.

The MCMC[M] code returns the DEM, among any user-defined number
of possible DEMs that are 
generated based on the MCMC method, that best reproduces the observed
fluxes.  
%For the reference composition we use the solar compilation of
%\citet{Grevesse98}. 
As noted above, one product of our MCMC[M] method is the statistical
uncertainties in $\Phi(T)$ evaluated at the 68\% confidence level. One
has to bear in mind that 
the error bars for consecutive bins in DEMs are not independent,
owing to the inevitable cross-talk that arises as a result of the line
contribution functions themselves stretching over several bins.
While the DEM method imposes a degree of local smoothing, dictated by
the number of diagnostic lines with significant emission in each
temperature bin \citep[see][ for details]{Kashyap98}, this smoothing is
not always sufficient to attenuate ``high frequency oscillations'' in
the solution---the artifact whereby overly high values of the DEM in a
given temperature bin can be compensated for by a commensurately 
lower value in adjacent bins.

Our final DEMs are corrected for the distances of the stars: we have adopted
$D$=14.9\,pc \citep{Guirado97} and $D$=48\,pc \citep{deBruijne01} for
AB Dor and V471 Tau respectively.

\subsection{Element Abundances}
\label{s:ele_abun}
Once the DEM, $\Phi(T)$, has been established, we can evaluate the
abundances of any elements for which we have lines with measured
fluxes.  The abundance is then a simple scaling factor in the integral
equation \ref{e:flux}, and is given by the value this factor has to
assume in order to match the measured flux.  In this way, we derived
values for the coronal abundances of the elements O, Ne, Mg, Si, S
and Fe, though only upper limits were obtained for the abundances of S for V471 Tau owing to the lack of signal in both He-like and
H-like features. According to previous works the photospheric abundances for AB Dor and V471 Tau are 
similar to the solar ones (see \S\ref{s:abdor} and \S\ref{s:v471tau}). Therefore, 
we will use the  \citet{Grevesse98} solar photospheric composition but will refer to 
it as stellar photospheric values. Note that we have used the new solar photospheric oxygen abundance, O/H = 8.69, derived by \citet{AllendePrieto01}.

We also used the temperature-insensitive abundance ratio diagnostics
of \cite{Drake04} as an additional check on our values obtained using
the DEM.  These are ratios of formed by combining two sets of lines of
two different elements, constructed such that the combined emissivity
curves of each set have essentially the same temperature
dependence. The resulting ratio of measured line fluxes then yields
directly the ratio of the abundances of the relevant elements,
independent from the atmospheric temperature structure. A similar,
though less sophisticated, method has been previously used in solar
coronal observations \citep[see, e.g.][]{Feldman92,Feldman00} and in
stellar EUVE observations \citep[][]{Drake95,Drake97}.

\section{Results and Discussion}
\label{s:results}

In order to verify the propriety of our DEM and abundance techniques, we
have compared observed and modelled line fluxes vs ionic species in
Fig~\ref{f:obs_pre} (upper panel) and vs $T_{max}$ (lower panel) for
the four line sets and for the final DEMs and abundances.  Open
symbols and filled symbols show the values for AB Dor and V471 Tau
respectively.  The great majority of the 
predicted line fluxes based on our final models are
within 10~\%\ or so of the observed values, with only four
lines being discrepant by more than 20\% .  We address these further
in \S\ref{s:structure} below.

Fig~\ref{f:synsp_abdor} and Fig~\ref{f:synsp_v471tau} show the
observed and synthetic spectrum of AB Dor and V471 Tau. The synthetic
spectra were computed using line emissivities from the CHIANTI
database, and were obtained with the derived DEMs and abundances, for
the four line sets. We also computed synthetic spectra using
emissivities from the APED database \citep{Smith01}, and the DEM and abundances derived from line set
1 for comparison.  For these simulations, we have use an electron
density of n$_{e}$=5\,10$^{11} cm^{-3}$ for AB~Dor and
n$_{e}$=2\,10$^{10} cm^{-3}$ for V471~Tau (Testa et al. 2004), with an
interstellar medium column density of n$_{H}$=1\,10$^{18} cm^{-2}$.
The exact density and very low ISM column makes no practical
difference to the synthetic spectra.  All the spectra are remarkably
similar with all the synthetic spectra showing good qualitative
agreement with observations for both stars.

\subsection{Temperature Structure}
\label{s:structure}

The reconstructed DEMs for AB~Dor and V471~Tau, together with
estimated 1$\sigma$ uncertainties, are shown in Fig.~\ref{f:dems}. The
DEMs are all quite similar in overall shape and
normalisation for the four sets of lines used.  A peak at
$\log\,T$[K]$\sim$7.0$\pm$0.1 is observed in all the reconstructions
independently of the set of lines used.  The DEMs for AB Dor also
exhibit secondary peaks at $\log T$[K]$\sim 6.6 \pm 0.1$ and $\log
T$[K]$\sim 7.4\pm 0.1$. The DEMs obtained by using the line sets with
no iron ions (sets 1 and 2), show sharper increase and decrease for
the lower and higher temperatures compared with ones obtained by
including iron lines (sets 3 and 4), which show more prominently
defined peaks.  Of these, it is interesting that the DEMs for both AB
Dor and V471 Tau corresponding to line set 4, containing all the ions,
require the most structure.

The DEMs for both stars show little evidence for substantial amount of
plasma at temperatures higher than $\log\,T$[K]=7.4.  Nevertheless,
the corona of AB~Dor seems to be, based on the DEMs, slightly more
active than that of V471~Tau.  The slopes in the derived DEMs to the
low temperature side of their maxima follow very approximately the
relation EM$\propto$T$^{2}$, which is steeper than the
EM$\propto$T$^{3/2}$ suggested for coronal loops with constant 
cross-section and uniform heating  \citep[e.g.;][]{Craig78,Jordan80,Peres00}. If the sharp peaks
observed in the DEMs are real, it would imply that the corona must
contain essentially isothermal structures \citep[see
also][]{Drake00,Scelsi04}.

Our results for AB Dor are comparable with those reported recently by
\citet{Sanz-Forcada03} based on XMM-{\it{Newton}} and {\it{Chandra}}
observations. They obtained a three-peak DEM with a more smooth
increase for the lower temperatures, $\log\,T$[K]$<$6.6, and sharp
decrease for the higher temperatures, $\log\,T$[K]$>$7.4. The
\citet{Sanz-Forcada03} line-based DEM also shows a maximum at
$\log\,T$[K]$\sim$7.0$\pm$0.1. To our knowledge, there exist no other
DEM analysis of V471 Tau by other authors with which to compare our
temperature structure results.

How reliable are the structural features observed in the DEMs?
\citet{Kashyap98} demonstrated that solar quiet and active region DEMs
are {\em not stable} to changes in the lines used in the
analysis. This occurs because of a breakdown in the assumptions of
optically thin plasma analysis, or (or in addition to) because of
uncertainties in the underlying atomic data that are not included in the
analysis.  We note in passing that measurements of the \ion{O}{8}
and \ion{Ne}{10} Ly$\alpha$/Ly$\beta$ ratios for AB Dor are in perfect
agreement with the optically thin theoretical values (Testa et al., in
preparation), so that we do not expect photon loss through, e.g.,
resonance scattering to be important.

The principal lines used in our analysis are simple H-like and 
He-like transitions for which atomic data should be reliable.  The
exceptions in line sets 3 and 4 are the Fe ions.  \citet{Doron02} and
\citet{Gu03} reported new Fe L-shell calculations indicating that rate
coefficients for all n$\rightarrow$2 (3$<$n$<$5) transitions may
differ from earlier calculations up to 50\%.  Moreover, these authors
emphasised the importance of including in the level populations the
indirect processes of radiative recombination, dielectronic
recombination, and resonance excitation involving neighbouring charge
states.  It seems likely that problems associated either with the Fe
line emissivities themselves, or with the underlying ionization
balance, might be to blame for the differences in the temperature
structure obtained from the different line sets.  Regarding the
ionization balance, we also note that dielectronic recombination (DR) tends
to dominate the recombination rates for the Fe ions near their
temperatures of peak population.  In contrast, for H-like ions, DR is
not an issue and one seems justified in having more faith in the
populations of these ions. The case of He-like ions has been discussed recently 
by \citet{Smith01} based on the example of O~VII.
Unlike the excitation of the {(1s.2p) 3S$_1$} and {(1s.2p) P$_{1,2}$} 
levels that form the upper levels of the prominent forbidden and 
intercombination lines, respectively, and which tend to be dominated by 
dielectronic recombination and cascades,
the {(1s.2p) 1P$_1$} upper level of the resonance line is excited 
almost entirely by collisions.  Thus, the He-like resonance lines should 
also be quite accurately represented in recent theoretical models and 
should provide reliable diagnostics.

\begin{figure}
%\begin{turn}{90}
\plotone{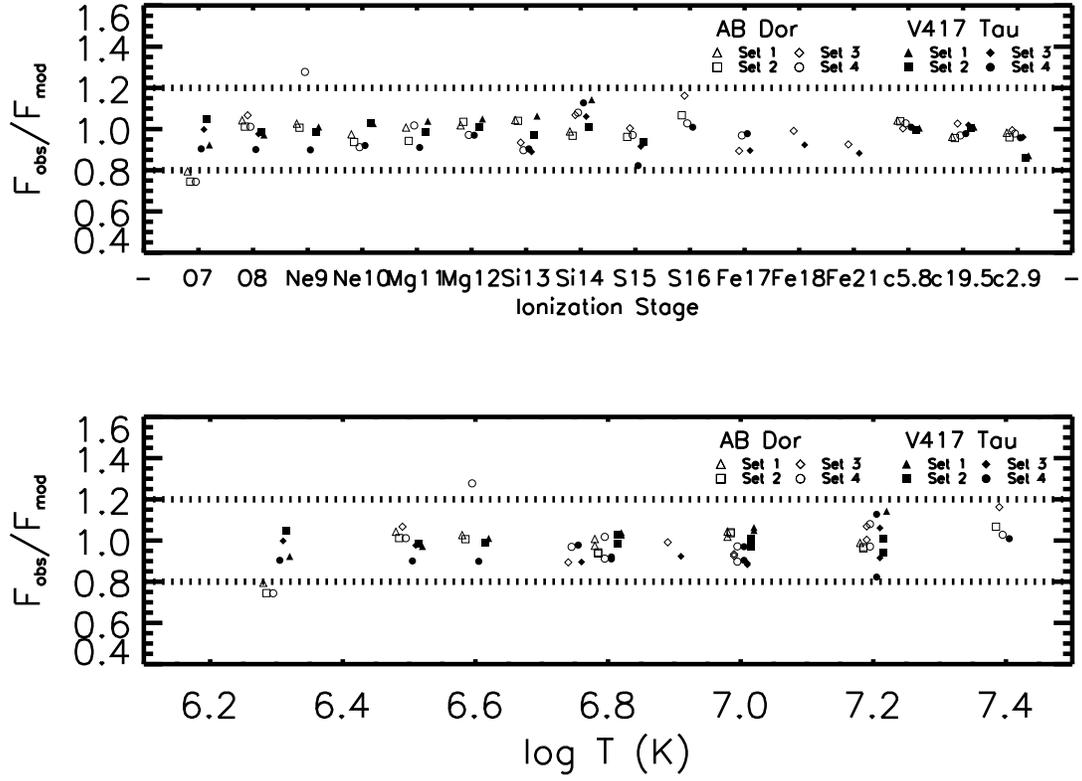}
%\end{turn}
\caption{Comparison of observed and modelled line fluxes vs ionic
species (upper panel) and vs $T_{max}$ (lower panel) for the four
sets. Open symbols and filled symbols show the values for AB Dor and
V471 Tau respectively. The dashed lines represent 1-$\sigma$
deviation.}
\label{f:obs_pre}
\end{figure}

As we noted earlier in \S\ref{s:results}, a small fraction of our
predicted line fluxes based on final DEMs and abundances differ to
those observed by $20$~\%\ or slightly more (Fig~\ref{f:obs_pre}).
Conspicuous among these 
are the O~VII lines in AB~Dor.  This small discrepancy is likely due
to the lack of strong temperature constraints for the cooler
temperatures---a similar, though smaller, effect is discernible at the
high temperature end from S~XVI.  One other conspicuous point in 
Fig~\ref{f:obs_pre} belongs to Ne~IX in AB~Dor for line set 4 that
includes Fe lines in addition to the H-like and He-like ions.  Since
this Ne~IX predicted flux is the only discrepant point among the
three line lists in which it appears, we conclude that there is no DEM
solution that can accommodate both the Fe and other lines
simultaneously to within their statistical uncertainties.  In essence, the
Fe~XVII and Fe~XVIII lines, that are formed at temperatures similar to
those of Ne~IX and Ne~X, are pulling the DEM to a solution that is
unfavourable to the Ne IX line.  

We have argued above that atomic data uncertainties are a likely
culprit for this situation.  However, another possibility is that the
coronal abundances are non-uniform as a function of temperature. This
would also be seen in the differences between DEMs derived by using set 1
and set 2, in which we use Mg and Ne ratios for the low temperature
range, and set 3 in which we swapped those ratios with Fe ones. The
latter results in a much lower DEM at $\log T\sim 6.6$, near where Fe~XVII
ions reach maximum population.  To retain a similar DEM to those
obtained from line sets 1 and 2 at this
temperature would require an {\em increase} in the Fe abundance at this
temperature relative to the value at higher temperatures.
Equally, similar agreement of the DEM's from line sets 1, 2 and 3
might be achieved by a {\em decrease} in the Ne abundance toward
higher temperatures.  These arguments are qualitative and we have
insufficient spectroscopic diagnostic information to investigate
abundance variations with temperature in a rigorous fashion.  We note,
however, that such abundance variations might not be unreasonable in
the light of a recent survey of coronal densities by \citet{Testa04} that clearly demonstrates that the cooler
$T\sim 10^6$~K corona resides in quite different structures to the
hotter, $T\sim 10^7$~K plasma that appears to be universally
characterised by much higher pressures. \citet{Feldman00} found that for the solar corona the FIP effect is larger for $T\sim 10^{5.9}$~K than at $T\sim 10^{6.15}$~K. 

Regardless of the true cause of the differences between the DEMs obtained
from different line sets, we can conclude that {\em small-scale structure
obtained in coronal DEMs is not reliable}: such structure 
sensitive to the particular diagnostics
chosen and does not necessarily reflect the true underlying
temperature structure.  This conclusion echoes the statements made
earlier by \citet{Kashyap98}.

The DEMs of V471~Tau are more similar among the different line sets
than those of AB~Dor.  However, despite of their quite different
evolutionary histories, we conclude that their coronae show a strong
similarity in temperature structure.

\subsection{Coronal Abundances}

Our results for the element abundances of AB Dor and V471 Tau are
summarized in Table~\ref{t:abund1}. We only list statistical
uncertainties here.  There are of course other uncertainties in the
atomic data used for the analysis, and in the calibration of the {\it
Chandra} instruments.  Atomic data uncertainties are probably of order
30\%\ for Fe lines \citep[e.g.,][]{Drake95}, but are probably slightly
less than this for H-like and He-like lines.  The final uncertainties
in our derived abundances are difficult to assess rigorously, but 
are likely of order 0.1~dex.

The abundances derived are quite consistent for the different line sets
considered: the abundances for each element agree even within the
statistical errors except for oxygen in set 3 of AB~Dor.  We also
notice that the abundances for oxygen and iron for both AB Dor and
V471 Tau derived from set 3 are slightly higher than those derived
from the other sets, an effect due to differences in the temperature
structures obtained (see \S\ref{s:structure}).

Table~\ref{t:abund2} shows the abundance ratio, from He-like and H-like ions of light elements, that are relatively insensitive to temperature (see \S\ref{s:ele_abun}). Those values are in agreement with the ones derived from the DEMs (Table~\ref{t:abund1}), with the exception of the ratios containing iron. Atomic physics uncertainties in the iron ions might affect the values obtained for the ratios [Ne/Fe] and [Mg/Fe].

In Fig.~\ref{f:abund} we have plotted the derived abundances, relative
to the stellar photospheric values, in order of the element FIP.
AB~Dor and V471~Tau clearly share similarities in coronal composition:
both show the same general pattern characterised by a depletion of low
FIP elements by a factor of $\sim 3$, evidence for slightly smaller
depletions of intermediate FIP elements S and O, and a clear
enhancement of Ne relative to elements of lower FIP.  This pattern is
similar to, though less extreme than, those seen in other active stars
and RS~CVn-type binaries\citep[see, e.g., reviews by][]{Drake03a,Audard03}.
That the abundances found here represent real metal depletions in the
coronae of both AB~Dor and V471~Tau is beyond doubt.  As discussed in
\S\ref{s:stars}, both stars have solar-like, or slightly super-solar,
metal abundances and this is irreconcilable with the much lower
coronal abundances of low FIP elements.

In Table~\ref{t:abund_liter} we list the abundance results from
earlier studies of AB~Dor and V471~Tau
\citep{Mewe96,Ortolani98,Maggio00,Guedel01,Sanz-Forcada03,Still03},
together with our own. The values from \citet{Guedel01} 
are from the RGS results. Note that the values for S and Si are few times smaller 
than those obtained with EPIC from the same observations. 
The S value obtained with EPIC is comparable with our values and previous 
works on AB Dor. The results obtained from the literature are all compared in
Fig.~\ref{f:abund_liter}. It is important to remember that different
authors adopt different values for ``solar'' photospheric abundances:
for both the table and figure we have scaled the various abundance
results to the photospheric scale adopted here, based on the
solar mixture of \citet{Grevesse98}, which is similar to the stellar 
photospheric values derived for a few elements of the stars. Note that we have used the new solar photospheric oxygen abundance, O/H = 8.69, derived by \citet{AllendePrieto01}.

While there is good
agreement in the general trend of abundances as a function of FIP,
there are clearly some significant differences between the different
studies. The large spread in results from the different studies 
appear to be a result of modelling uncertainties and possibly to improvements 
of the atomic database. This conclusion is reinforced by the much more favourable 
comparison of the results of the recent {\it Chandra} and {\it
XMM-Newton} studies of AB~Dor---this work compared with that of
\citet{Guedel01} and \citet{Sanz-Forcada03}.

\citet{Audard03b}, \citet{Huenemoerder03} and \citet{Sanz-Forcada03} have recently reported, for V824 Ara, AR LAc and AB Dor respectively,
higher values for abundances of Al, Ca, and Na than for elements with
slightly higher FIP suggesting a schism in the
strictly inverse FIP effect scheme.  In particular,
\citet{Sanz-Forcada03} obtained different values of Ca abundance for
AB~Dor, including a very high value of [Ca/H]=0.62 derived from the high resolution RGS spectrum. Indeed, Figure~12 of
\citet{Sanz-Forcada03} suggests that the inverse FIP effect in AB~Dor
undergoes a palpable reversal toward the lowest FIP elements.  A
similar effect was suggested by \citet{Huenemoerder03} based on high
resolution {\it Chandra} spectra of AR~Lac.  That there might be a
difference in abundance behaviour among the low FIP elements has
already been suggested on the basis of solar measurements, in which
the lowest FIP elements, such as Na and Al, appear to be enhanced
relative to elements such as Si and Mg \citep{Feldman92}.

Although we have not been able to obtain definitive measurements of
lines of Na, Al and Ca in our spectra of AB~Dor (none were detected
with any degree of significance), we have estimated approximate upper
limits to the abundances of these elements through comparison of
observed and synthetic spectra.  Fig~\ref{f:low_fip_synt} illustrates
the observed AB~Dor MEG spectrum and two synthetic spectra: one
computed using the abundance derived for Ca by \citet{Sanz-Forcada03}
and the other one assuming Al, Ca and Na have coronal abundances
similar to Mg.  The former produce extremely strong \ion{Ca}{19},
\ion{Al}{12} and \ion{Na}{11} lines which are clearly not seen in the
observed spectrum.  The synthetic spectrum computed assuming relative coronal
abundances equal to Mg for Al, Ca and Na instead provide good
approximations to the observed spectrum.  We conclude that the coronal
abundances of Al, Ca and Na in AB~Dor are similar to, or lower than,
those of the other low FIP elements Mg, Si and Fe.

The coronal abundances we have derived for V471 Tau are a factor of
two or less lower (factor three for Ne) than those reported by \citet{Still03}. This might be due
to artifacts of the analysis of the low resolution ASCA pulse height
spectra, though other explanations are possible.  There were three
modest flare events during the ASCA observation that were not filtered
from the analysis of \citet{Still03}.  It appears likely that in some
flares on active stars the abundances of low FIP elements are enhanced
toward their pristine photospheric values, a result tentatively
interpreted as chromospheric evaporation \citep[e.g. reviews of][]{Audard03,Drake03a}.  Flares, then, might temporarily raise the
abundances of the dominant plasma at a given time, though the ASCA
flares were rather small for this to present an immediately convincing
explanation.

Alternatively, it is possible that the abundance patterns simply vary
with time, according to the characteristics of dominant and relatively
quiescent active regions, and at the time of the earlier ASCA
observation the abundances were simply higher.  Circumstantial
evidence against such an explanation can be found in the abundance
analysis of the active RS~CVn-type binary HR~1099, for which
independent analyses of {\it Chandra} and {\it XMM-Newton}
observations obtained at different times yielded very similar values
for quiescent coronal abundances \citep{Brinkman01,Drake01}.
Moreover, \citet{Drake03b} has found similar abundances in the coronae
of HR~1099 and its ``coronal twin'' the active late-type component of
the Algol binary.  

That our results for AB~Dor and V471~Tau---coronal
twins in many respects---are so similar reinforces the conclusion that
coronal abundances are dependent largely on stellar parameters and do
not vary substantially during times of relative coronal quiescence. It also 
points out that tidal interactions, which may have an influence on differential 
rotation and possibly on dynamo action, are not important for determinnig 
coronal structure (the magnetic interactions between the white dwarf and the K star 
components of V471~Tau does not seem to play an important role at least in the X-ray domain). However, 
despite 
the very similar coronal temperature structure and abundance patterns, the L$_X$ of 
V471 Tau is about twice higher than that for AB Dor (see Table~\ref{t:par}). 
This difference in brightness might be within the
stochastic variation range for active stars, but might also be hinting that 
tidal interaction could be contributing to the stellar coronae somehow.

%In summary, the ratio-based analysis applied to {\it{Chandra}} high
%resolution spectra of AB Dor and V471 Tau revleas a clear and pattern
%of the abundances versus FIP.  The observed anomalies and variation
%with FIP appear less extreme than those seen in very active stars,
%despite AB Dor and V471 Tau being at the limit between saturation and
%supersaturation regimes \citep[see, e.g.,][]{Pizzolato03,Jardine04}.
%Our analysis of the AB Dor HETG and V471 Tau LETG spectra show that
%magnesium, iron, silicon, sulfur, oxygen and neon are depleted
%relative to photospheric values. We have estimated upper limit for the
%coronal abundances of low FIP elements (Ca, Al and Na) which seems to
%be similar to the abundances derived for Mg. Nevertheless appears that
%both the DEMs and the abundances are independent on the line ratios
%used in the analysis.

\section{Conclusions}
\label{s:concl}

AB~Dor and V471~Tau represent a rare pair stars that have the same
rotational period (P$_{orb}\sim0\fd5$) and spectral type but quite
different evolutionary histories and ages($\sim$70\,Myr vs
$\sim$600\,Myr).  As such, a comparison between their coronal
properties could provide an illuminating glimpse of any fundamental
underlying differences in their magnetic dynamos and activity.  Based
on an analysis of high resolution {\it Chandra} X-ray spectra of these
stars we draw the following conclusions.

\begin{enumerate}
\item Both the coronal temperature structure and element abundances
are remarkably similar. Element abundances are characterized by an
inverse-FIP type effect in which the abundances of high FIP elements
are enhanced relative to those of low FIP elements.  Both coronae are
depleted in low FIP elements relative to their photospheres by a
factor of $\sim 3$.  

\item The observed abundance anomalies are less extreme than those
typically seen in very active stars, despite AB Dor and V471 Tau lying
at the boundary between coronal saturation and supersaturation regimes
\citep{Pizzolato03,Jardine04}.

\item There is no evidence for any difference in the abundances of the
very low FIP ($\sim5$-6~eV) elements Na, Al and Ca as compared to the low FIP
($\sim 7.5$-8~eV) elements Si, Mg and Fe, as some earlier work has
suggested.    

\item Abundances derived using our method in which the DEM is obtained
from a combination of the ratios of a few well-selected prominent
spectral lines and continuum measurements are robust and independent
of both the exact set of lines used to define the DEM and of the
details of its temperature structure.  

\item Small-scale structure is found in the DEMs obtained based on
different sets of spectral lines.  Such structural details are most likely spurious artifacts arising from either
errors in assumed ion populations or spectral line excitation rates,
or non-uniform abundances in different coronal structures that
contribute significantly to the observed X-ray emission.  {\em Coronal
DEMs are strongly dependent on the input data used to derive them}, as
was emphasised earlier in the solar study of \citet{Kashyap00}.

\item Based on the very similar coronal characteristics of AB~Dor and
V471~Tau we conclude that there is no observable difference between
the coronal activity of youth and high primordial angular momentum and
that of tidal spin-up of an older star. In other words, the exact
evolutionary state has little effect on the corona, and the parameters
that dominate coronal structure and composition are rotation rate and
spectral type.  The latter determines of course the parameters of the
convection zone, where the magnetic fields sustaining coronal activity
are believed to originate, while rotation rate appears to be the
dominant parameter in the amplification of these magnetic fields.

\item We also note that despite the binarity of V471 Tau it does not show any significant difference in thermal structure and composition with respect to AB~Dor. This would mean that tidal interactions, which may have an influence on differential rotation and possibly on dynamo action, are not important for determining coronal structure. The magnetic interactions between the white dwarf and the K star components of V471~Tau does not seem to play an important role at least in the X-ray domain. The latter conclusion was previously suggested in an analysis based on radio observations \citep{Lim96}. Nevertheless, the observed difference in brightness between AB Dor and V471 Tau (twice higher) might be hinting that tidal interaction could be contributing to the stellar coronae somehow.

\end{enumerate}

\acknowledgments

DGA and WB were supported by {\it{Chandra}} grants GO1-2006X and
 GO1-2012X. LL was supported by NASA AISRP contract NAG5-9322; we
 thank this program for providing financial assistance for the
 development of the PINTofALE package. We also thank the CHIANTI
 project for making publicly available the results of their
 substantial effort in assembling atomic data useful for coronal
 plasma analysis.  JJD and VK were supported by NASA contract
 NAS8-39073 to the {\it{Chandra}}. We also thank the referee, Manuel 
 G{\" u}del, for helpful comments on the manuscript.

\clearpage

%%%%%%%%%%%%%%%%%%%%%%%%%%%%%%%%%%%%%%%%%%%%%%%%%%%%%%

\newpage\section*{Tables}

\begin{deluxetable}{lrcccccccc}
\tabletypesize{\scriptsize}
\tablecaption{Summary of Stellar Parameters. \label{t:par}}
\tablewidth{0pt}
\tablehead{ \colhead{} & \colhead{Spectral} & 
\colhead{Period} & \colhead{Distance} 
%& \colhead{T$_eff$}& \colhead{Radius} & \colhead{Mass} 
& \multicolumn{3}{c}{Active Component} & \colhead{i} & \colhead{$v$\,sin$i$} & \colhead{L$_x$}\\
\colhead{Star} & \colhead{Type} & \colhead{[d]} & \colhead{[pc]} 
& \colhead{T$_{eff}$[K]}& \colhead{$R_\odot$} & \colhead{$M_\odot$}  & \colhead{[deg]} & \colhead{[km\,s$^{-1}$]} & \colhead{[10$^{30}$\,erg\,s$^{-1}$]}} 
\startdata
AB Dor	& K0-2\,V\tablenotemark{a} &  0.51\tablenotemark{b} & 14.9\tablenotemark{c} & $5250$\tablenotemark{d} &0.86\tablenotemark{e} & 0.76\tablenotemark{c} & 60\tablenotemark{f} & 93\tablenotemark{e}&0.70\tablenotemark{l} \\
V471\,Tau & K2\,V\,+\,WD\tablenotemark{g} &  0.52\tablenotemark{h} & 48\tablenotemark{i} & $5040$\tablenotemark{j} &0.96\tablenotemark{j} &0.93\tablenotemark{j}&77\tablenotemark{j}&91\tablenotemark{k}&1.58\tablenotemark{l} \\
\enddata
\tablenotetext{a}{Parameters of the component that dominates X-ray activity: AB Dor and the K2\,V component of V471\,Tau.}
\tablenotetext{a}{\citet{Collier-Cameron89}}
\tablenotetext{b}{\citet{Pakull81}}
\tablenotetext{c}{\citet{Guirado97}}
\tablenotetext{d}{\citet{Johnson66}}
\tablenotetext{e}{\citet{Maggio00}}
\tablenotetext{f}{\citet{Kuerster94}}
\tablenotetext{g}{\citet{Rucinski81}}
\tablenotetext{h}{\citet{Guinan01}}
\tablenotetext{i}{\citet{deBruijne01}}
\tablenotetext{j}{\citet{OBrien01}, who notes that these values are 
for $i=77\degr$ and are somewhat sensitive to the
inclination adopted; in particular the mass and radius are larger for
smaller inclinations.}
\tablenotetext{k}{\citet{Ramseyer95}}
\tablenotetext{l}{Derived from our {\it{Chandra}} observations.}
\end{deluxetable}

\begin{deluxetable}{lrrrr}
\tabletypesize{\scriptsize}
\tablecaption{Summary {\it{Chandra}} Observations.\label{t:obs}}
\tablewidth{0pt}
\tablehead{  &\colhead{ObsID} & \colhead{Instrument} & \colhead{Start} & \colhead{Exp}\\
\colhead{Star} &\colhead{}& &\colhead{[UT]} & \colhead{[ks]}
} 
\startdata
AB Dor	& 0016 & HETG\tablenotemark{a}+ACIS-S&1999-10-09T11:21:09 & 52.3 \\
V471\,Tau &  2523 &LETG\tablenotemark{b}+HRC-S & 2002-01-24T22:30:03 & 87.5 \\
\enddata
\tablenotetext{a}{\citet{Weisskopf02}}
\tablenotetext{b}{\citet{Predehl97}}

\end{deluxetable}

\begin{deluxetable}{lllcrrl}
\tabletypesize{\scriptsize}
\tablecaption{Identification and fluxes for spectral lines, observed on AB Dor and V471 Tau, used in this analysis.\label{t:flx}}
\tablehead{
\colhead{$\lambda_{\rm obs}$} & 
\colhead{$\lambda_{\rm pred}$} &
\colhead{Ion} & 
\colhead{$log\,T_{\rm max}$\tablenotemark{a}} & 
\colhead{AB Dor}  & 
\colhead{V471 Tau} & 
\colhead{Transition} \\
\colhead{(\AA )} &
\colhead{(\AA )} &
\colhead{} &
\colhead{(K)} & 
\colhead{(10$^{-5}$ ph/cm$^2$/s)} & 
\colhead{(10$^{-5}$ ph/cm$^2$/s)} & 
\colhead{upper $\rightarrow$ lower}
}

\startdata
 4.722 &  4.727 & S  XVI  & 7.40 & $  1.56 \pm 0.59$ & $ 0.33 \pm 0.72$  &(2p) 2P$_{3/2}$ $\rightarrow$ (1s) 2S$_{1/2}$ \nl
\nodata & 4.733 & S  XVI  & 7.40 & \nodata	     & \nodata         & (2p) 2P$_{1/2}$ $\rightarrow$ (1s) 2S$_{1/2}$ \nl
 5.032 &  5.039 & S  XV   & 7.20 & $  2.26 \pm 0.75$ & $ 0.62 \pm 0.70$  &(1s.2p) 1P$_{1}$ $\rightarrow$ (1s2) 1S$_{0}$ \nl
 6.177 &  6.180 & Si XIV  & 7.20 & $  4.80 \pm 0.38$ & $ 0.59 \pm 0.28$  &(2p) 2P$_{3/2}$ $\rightarrow$ (1s) 2S$_{1/2}$ \nl
\nodata & 6.186 & Si XIV  & 7.20 & \nodata	     & \nodata         & (2p) 2P$_{1/2}$ $\rightarrow$ (1s) 2S$_{1/2}$ \nl
 6.647 &  6.648 & Si XIII & 7.00 & $  4.79 \pm 0.40$ & $ 0.85 \pm 0.34$  &(1s.2p) 1P$_{1}$ $\rightarrow$ (1s2) 1S$_{0}$ \nl
 8.422 &  8.425 & Mg XII  & 7.00 & $  5.95 \pm 0.39$ & $ 0.88 \pm 0.35$  &(2p) 2P$_{1/2}$ $\rightarrow$ (1s) 2S$_{1/2}$ \nl
\nodata & 8.419 & Mg XII  & 7.00 & \nodata	     & \nodata         & (2p) 2P$_{3/2}$ $\rightarrow$ (1s) 2S$_{1/2}$ \nl
 9.173 &  9.169 & Mg XI   & 6.80 & $  3.71 \pm 0.39$ & $ 0.61 \pm 0.42$  &(1s.2p) 1P$_{1}$ $\rightarrow$ (1s2) 1S$_{0}$ \nl
10.028 & 10.029 & Na XI   & 6.90 & $  2.89 \pm 0.65$ & $ 0.73 \pm 0.30$  &(2p) 2P$_{1/2}$ $\rightarrow$ (1s) 2S$_{1/2}$ \nl
\nodata & 10.023 & Na XI & 6.90  & \nodata	     & \nodata         & (2p) 2P$_{3/2}$ $\rightarrow$ (1s) 2S$_{1/2}$ \nl
10.998 & 11.003 & Na X    & 6.70 & $  8.56 \pm 0.85$ & $ 1.23 \pm 0.36$  &(1s.2p) 1P$_{1}$ $\rightarrow$ (1s2) 1S$_{0}$ \nl
12.133 & 12.132 & Ne X    & 6.80 & $ 68.06 \pm 1.80$ & $ 8.00 \pm 0.77$  &(2p) 2P$_{3/2}$ $\rightarrow$ (1s) 2S$_{1/2}$ \nl
\nodata & 12.137 & Ne X	 & 6.80  & \nodata	     & \nodata         & (2p) 2P$_{1/2}$ $\rightarrow$ (1s) 2S$_{1/2}$ \nl
12.283 & 12.285 & Fe XXI  & 7.00 & $ 14.33 \pm 1.33$ & $ 1.74 \pm 0.52$  &(2s2.2p3d) 3D$_{1}$ $\rightarrow$ (2s2.2p2) 3P$_{0}$ \nl
13.433 & 13.447 & Ne IX   & 6.60 & $ 32.96 \pm 1.79$ & $ 3.02 \pm 0.53$  &(1s.2p) 1P$_{1}$ $\rightarrow$ (1s2) 1S$_{0}$ \nl
13.508 & 13.504 & Fe XIX  & 6.90 & $ 13.87 \pm 1.62$ & $ 3.58 \pm 0.57$  &(2p3(2P).3d) 1D$_{2}$ $\rightarrow$ (2s2.2p4) 3P$_{2}$ \nl
14.208 & 14.208 & Fe XVIII& 6.90 & $ 21.24 \pm 1.62$ & $ 2.29 \pm 0.54$  &(2p4(1D).3d) 2P$_{3/2}$ $\rightarrow$ (2s2.2p5) 2P$_{3/2}$ \nl
\nodata &14.203 & Fe XVIII& 6.90 & \nodata	     & \nodata         & (2p4(1D).3d) 2D$_{5/2}$ $\rightarrow$ (2s2.2p5) 2P$_{3/2}$ \nl
14.263 & 14.267 & Fe XVIII& 6.90 & $  5.69 \pm 1.09$ & $ 1.27 \pm 0.47$  &(2p4(1D).3d) 2F$_{5/2}$ $\rightarrow$ (2s2.2p5) 2P$_{3/2}$ \nl
15.013 & 15.015 & Fe XVII & 6.75 & $ 36.18 \pm 2.01$ & $ 5.23 \pm 0.62$  &(2p5.3d) 1P$_{1}$ $\rightarrow$ (2p6) 1S$_{0}$ \nl
16.008 & 16.007 & O  VIII & 6.50 & $ 33.87 \pm 2.13$ & $ 3.56 \pm 0.60$  &(3p) 2P$_{1/2}$ $\rightarrow$ (1s) 2S$_{1/2}$ \nl
\nodata &16.006 & O VIII  & 6.50 & \nodata	     & \nodata         & (3p) 2P$_{3/2}$ $\rightarrow$ (1s) 2S$_{1/2}$ \nl
18.973 & 18.973 & O  VIII & 6.50 & $154.21 \pm 6.18$ & $14.32 \pm 1.05$  &(2p) 2P$_{1/2}$ $\rightarrow$ (1s) 2S$_{1/2}$ \nl
\nodata &18.967 & O VIII  & 6.50 & \nodata	     & \	         & (2p) 2P$_{3/2}$ $\rightarrow$ (1s) 2S$_{1/2}$ \nl
21.607 & 21.602 & O  VII  & 6.30 & $ 17.49 \pm 4.16$ & $ 2.71 \pm 0.84$  &(1s.2p) 1P$_{1}$ $\rightarrow$ (1s2) 1S$_{0}$ \nl
24.777 & 24.779 & N  VII  & 6.30 & $ 23.01 \pm 4.70$ & $ 4.84 \pm 1.07$  &(2p) 2P$_{3/2}$ $\rightarrow$ (1s) 2S$_{1/2}$ \nl
\nodata &24.785 & N VII	  & 6.30 & \nodata          & \nodata            & (2p) 2P$_{1/2}$ $\rightarrow$ (1s) 2S$_{1/2}$ \nl
\enddata
\tablenotetext{a}{The temperature at which the function $G_{ij}(T)$ peaks.}
\end{deluxetable}

% refs from chianti for collisional excitation:
% C VI Aggarwal,K.M., Kingston,A.E., 1991, J. Phys. B, 24, 4583
% N VII Aggarwal,K.M., Kingston,A.E., ??
% O VIII ??
% Ne X Aggarwal, K. M., and Kingston, A. E., 1991, Phys. Scripta, 44, 517
\newpage

%=======================================
%=======================================
%=======================================
%===========    CORRECTED   ============
%=======================================
%=======================================
%=======================================
\begin{table}
\caption{Coronal abundances obtained from an abundance-independent DEM-reconstruction for AB Dor and V471 Tau, relative to stellar photospheric values. Note: The abundances have been calculated from four different set of ions for each object.\label{t:abund1}}
%\tiny{
\scriptsize{
\begin{center}
\begin{tabular}{lrcrrrr}
\hline  \hline \\
Element\tablenotemark{a}& 
FIP\tablenotemark{b}&
X/H\tablenotemark{c}&
AB Dor\tablenotemark{d}&
AB Dor\tablenotemark{e}&
AB Dor\tablenotemark{f}&
AB Dor\tablenotemark{g}\\
\hline \\
$[$Na/H$]$&5.14 &6.33&   $<$0.54\tablenotemark{h}         &   $<$0.54\tablenotemark{h}&   $<$0.54\tablenotemark{h}&   $<$0.51\tablenotemark{h}\\
$[$Al/H$]$&5.98 &6.47&   $<$0.54\tablenotemark{h}         &   $<$0.54\tablenotemark{h}&   $<$0.54\tablenotemark{h}&   $<$0.51\tablenotemark{h}\\
$[$Ca/H$]$&6.11 &6.36&   $<$0.54\tablenotemark{h}         &   $<$0.54\tablenotemark{h}&   $<$0.54\tablenotemark{h}&   $<$0.51\tablenotemark{h}\\
$[$Mg/H$]$&7.65 &7.58&  -0.56$\pm$   0.03&  -0.56$\pm$   0.03&  -0.61$\pm$   0.03&  -0.60$\pm$   0.03\\
$[$Fe/H$]$&7.87 &7.50&  -0.57$\pm$   0.04&  -0.56$\pm$   0.04&  -0.47$\pm$   0.03&  -0.54$\pm$   0.03\\
$[$Si/H$]$&8.15 &7.55&  -0.50$\pm$   0.03&  -0.49$\pm$   0.04&  -0.53$\pm$   0.03&  -0.50$\pm$   0.03\\
$[$S/H$]$ &10.36&7.33&  -0.51$\pm$   0.11&  -0.49$\pm$   0.11&  -0.54$\pm$   0.11&  -0.47$\pm$   0.11\\
$[$O/H$]$ &13.62&8.69&  -0.22$\pm$   0.04&  -0.21$\pm$   0.04&  +0.07$\pm$   0.04&  -0.14$\pm$   0.04\\
$[$Ne/H$]$&21.56&8.08&  -0.10$\pm$   0.03&  -0.09$\pm$   0.03&  -0.03$\pm$   0.03&  -0.06$\pm$   0.03\\\\
\hline  \hline \\
Element\tablenotemark{a}& 
FIP\tablenotemark{b}&
X/H\tablenotemark{c}&
V471 Tau\tablenotemark{d}&
V471 Tau\tablenotemark{e}&
V471 Tau\tablenotemark{f}&
V471 Tau\tablenotemark{g}\\
\hline \\
$[$Na/H$]$&5.14 &6.33&  \nodata & \nodata  &  \nodata &  \nodata  \\
$[$Al/H$]$&5.98 &6.47&  \nodata & \nodata  &  \nodata &  \nodata  \\
$[$Ca/H$]$&6.11 &6.36&  \nodata & \nodata  &  \nodata &  \nodata  \\
$[$Mg/H$]$&7.65 &7.58&  -0.58$\pm$   0.17&  -0.56$\pm$   0.17&  -0.57$\pm$   0.16&  -0.58$\pm$   0.16 \\
$[$Fe/H$]$&7.87 &7.50&  -0.70$\pm$   0.12&  -0.66$\pm$   0.11&  -0.64$\pm$   0.09&  -0.66$\pm$   0.09 \\
$[$Si/H$]$&8.15 &7.55&  -0.53$\pm$   0.15&  -0.48$\pm$   0.15&  -0.48$\pm$   0.15&  -0.50$\pm$   0.15 \\
$[$S/H$]$ &10.36&7.33&  -0.37$\pm$   0.46&  -0.31$\pm$   0.45&  -0.30$\pm$   0.45&  -0.33$\pm$   0.45 \\
$[$O/H$]$ &13.62&8.69&  -0.32$\pm$   0.10&  -0.32$\pm$   0.10&  -0.25$\pm$   0.10&  -0.29$\pm$   0.09 \\
$[$Ne/H$]$&21.56&8.08&  -0.24$\pm$   0.09&  -0.24$\pm$   0.09&  -0.21$\pm$   0.09&  -0.23$\pm$   0.08 \\
\hline
\end{tabular}
\end{center}
\small{$^a$Abundances relative to stellar photospheric values.}\\
\small{$^b$First Ionization Potential in eV.}\\
\small{$^c$Assumed photospheric abundances expressed relative to H by number on a logarithm scale where H/H=12.}\\
\small{$^d$Abundances based on DEMs from O, Ne, Mg and Si.}\\
\small{$^e$Ditto from O, Ne, Mg, Si and S.}\\
\small{$^f$Ditto from O, Si, S, FeXVII, FeXVIII and FeXXI.}\\
\small{$^g$Ditto from O, Ne, Mg, Si, FeXVII, FeXVIII and FeXXI.}\\
\small{$^h$Upper limits based on synthetic spectra estimation.}\\
}
\end{table}

\begin{table}
\caption{AB Dor and V471 Tau Abundance Ratios using Temperature-Insensitive Diagnostics.\label{t:abund2}}
\begin{center}
\begin{tabular}{lrr}
\hline  \hline \\
Abundance Ratio&
AB Dor&
V471 Tau\\
\hline \\
$[$N/O$]$&   0.25$\pm$   0.09 & 0.53$\pm$   0.11 \\
$[$O/Ne$]$& -0.23$\pm$   0.11 &-0.19$\pm$   0.15 \\
$[$Ne/Mg$]$& 0.47$\pm$   0.04 & 0.32$\pm$   0.24 \\
$[$Ne/Fe$]$& 0.71$\pm$   0.03 & 0.57$\pm$   0.07 \\
$[$Mg/Si$]$&-0.09$\pm$   0.04 &-0.09$\pm$   0.22 \\
$[$Mg/Fe$]$&-0.18$\pm$   0.05 & 0.03$\pm$   0.31 \\
$[$Si/S$]$&  0.01$\pm$   0.12 &-0.18$\pm$   0.47 \\
\hline
\end{tabular}
\end{center}
\end{table}

\begin{table}
\caption{Comparison of coronal abundances from AB~Dor and V471~Tau obtained from this work and the literature. Note that all the coronal abundances are relative to stellar photospheric values.\label{t:abund_liter}}
%\tiny{
{
\begin{center}
\begin{tabular}{lrcrrrrrrrr}
\hline  \hline \\
Element\tablenotemark{a}& 
FIP\tablenotemark{b}&
X/H\tablenotemark{c}&
AB Dor\tablenotemark{d}&
AB Dor\tablenotemark{e}&
AB Dor\tablenotemark{f}&
AB Dor\tablenotemark{g}&
AB Dor\tablenotemark{h}&
AB Dor\tablenotemark{i}&
V471 Tau\tablenotemark{j}&
V471 Tau\tablenotemark{i}\\
\hline \\
$[$Na/H$]$&5.14 &6.33&  \nodata& \nodata  &\nodata &\nodata   &\nodata&-0.58\tablenotemark{k}&\nodata &\nodata\\
$[$Al/H$]$&5.98 &6.47&  \nodata& \nodata  &\nodata &-0.62     &-0.35  &-0.58\tablenotemark{k}&\nodata &\nodata\\
$[$Ca/H$]$&6.11 &6.36&  -0.31  & \nodata  &-0.74   &-0.99     & 0.41  &-0.58\tablenotemark{k}&\nodata &\nodata\\
$[$Ni/H$]$&7.63 &6.25&  -0.35  & -0.36	  &-0.33   &-0.64     &-0.18  &\nodata&\nodata &\nodata\\
$[$Mg/H$]$&7.64 &7.58&  -0.24  & -0.72	  &-0.57   &-0.58     &-0.39  &-0.58  &-0.06   &-0.57  \\
$[$Fe/H$]$&7.87 &7.50&  -0.47  & -0.73	  &-0.48   &-0.59     &-0.40  &-0.54  &-0.02   &-0.67  \\
$[$Si/H$]$&8.15 &7.55&  -0.46  & -0.80	  &-0.85   &-0.50     &-0.41  &-0.51  &-0.12   &-0.50  \\
$[$S/H$]$ &10.36&7.33&  -0.55  & -0.38	  &-1.52   &-0.66     &-0.46  &-0.50  &\nodata &-0.33  \\
$[$C/H$]$&11.26 &8.52&  \nodata& \nodata  &-0.30   &-0.32     &+0.00  &\nodata&\nodata &\nodata\\
$[$O/H$]$ &13.62&8.69&  -0.14  & -0.26	  &-0.16   &-0.31     &+0.16  &-0.13  &+0.19   &-0.30  \\
$[$N/H$]$&14.53 &7.92&  -0.32  & \nodata  &-0.15   &-0.21     &+0.22  &\nodata&\nodata &\nodata\\
$[$Ar/H$]$&15.76 &6.40& +0.16  & \nodata  &+0.09   &-0.30     &+0.19  &\nodata&\nodata &\nodata\\
$[$Ne/H$]$&21.56&8.08&  +0.09  & -0.20	  &+0.01   &-0.11     &+0.23  &-0.07  &0.39    &-0.23  \\\\
\hline
\end{tabular}
\end{center}
\small{$^a$Abundances relative to stellar photospheric values.}\\
\small{$^b$First Ionization Potential in eV.}\\
\small{$^c$Assumed photospheric abundances expressed relative to H by number on a logarithm scale where H/H=12.}\\
\small{$^d$\citet{Mewe96}}\\
\small{$^e${\citet{Ortolani98}}}\\
\small{$^f$\citet{Guedel01} based on RGS results}\\
\small{$^g$\citet{Sanz-Forcada03} based on 3-T analysis.}\\
\small{$^h$\citet{Sanz-Forcada03} based on DEM analysis.}\\
\small{$^i$This work}\\
\small{$^j$\citet{Still03}}\\
\small{$^k$Upper limits based on synthetic spectra estimation.}\\
}
\end{table}

\newpage

\newpage

\newpage

\newpage

\begin{figure}
%\begin{turn}{90}
\plotone{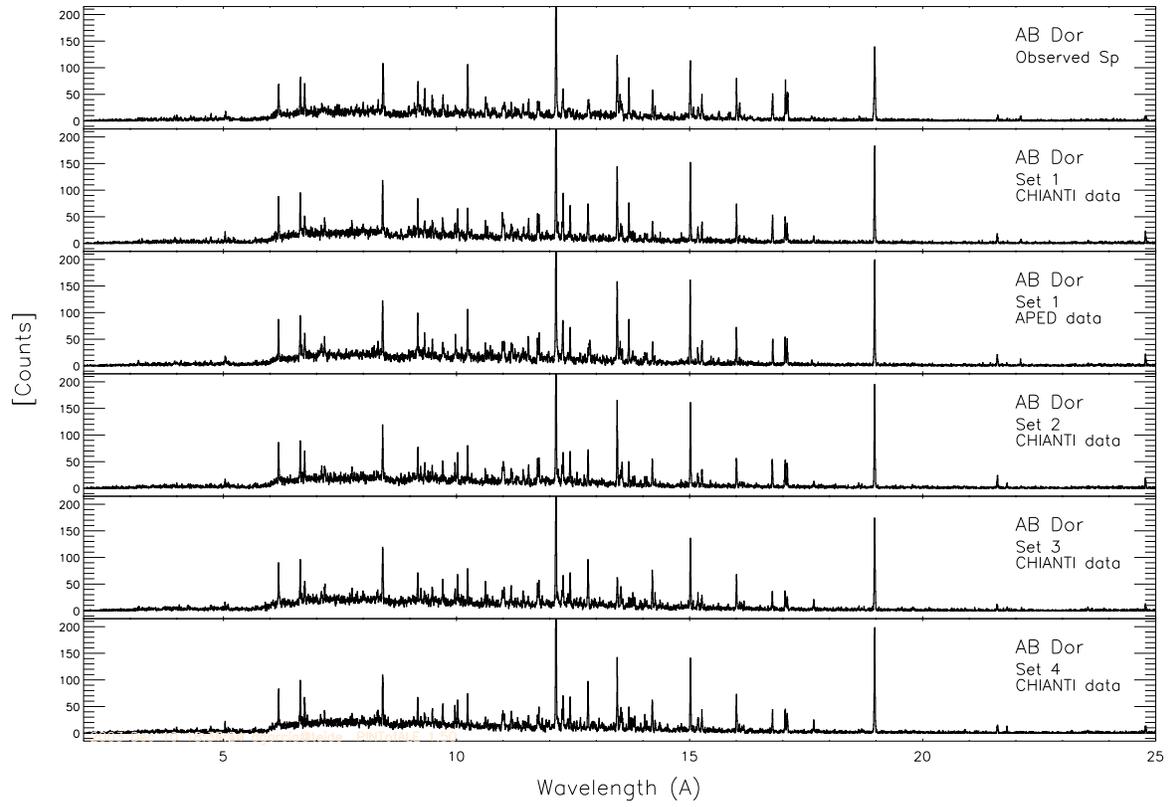}
%\end{turn}
\vspace{1cm}
\caption{Comparison between HETG+ACIS-S spectrum of AB Dor (top panel)
and the predicted spectrum derived from the four reconstructed DEMs.}
\label{f:synsp_abdor}
\end{figure}

\begin{figure}
%\begin{turn}{90}
\plotone{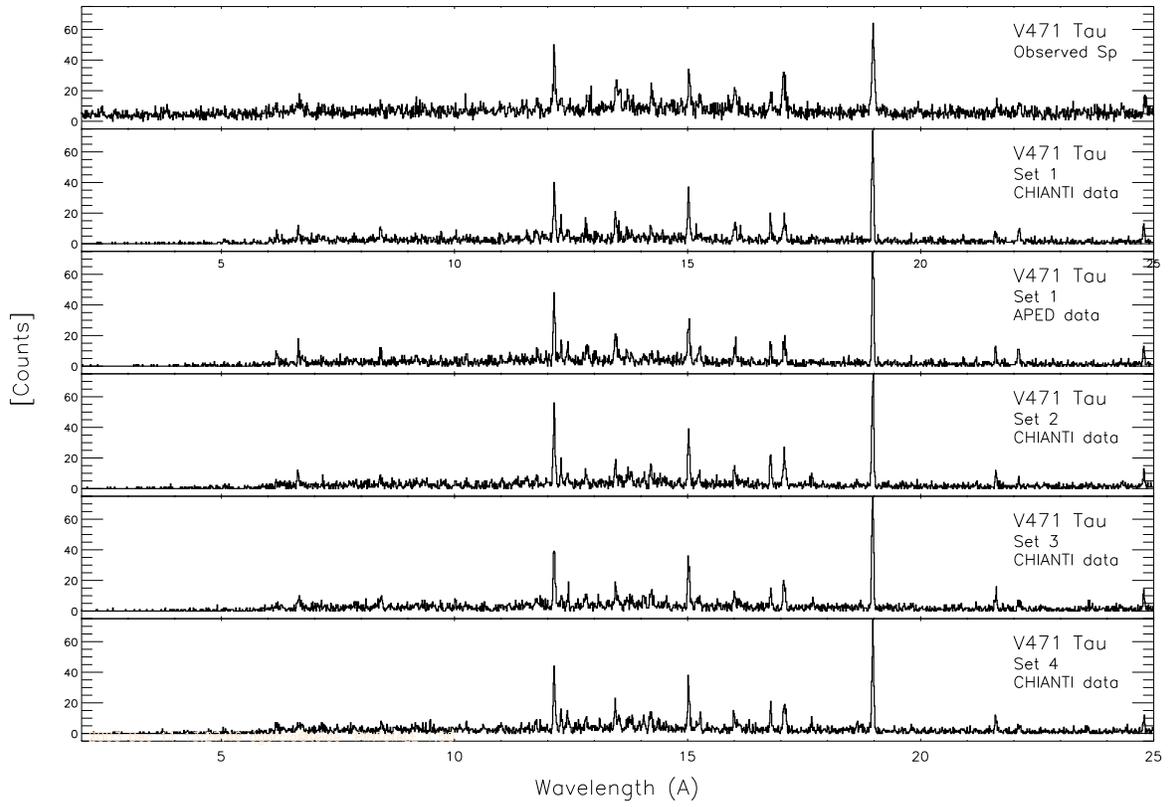}
%\end{turn}
\vspace{1cm}
\caption{Comparison between LETG+HRC-S spectrum of V471 Tau (top
panel) and the predicted spectrum derived from the four reconstructed
DEMs.}
\label{f:synsp_v471tau}
\end{figure}

\newpage

\begin{figure}
%\begin{turn}{90}
\plotone{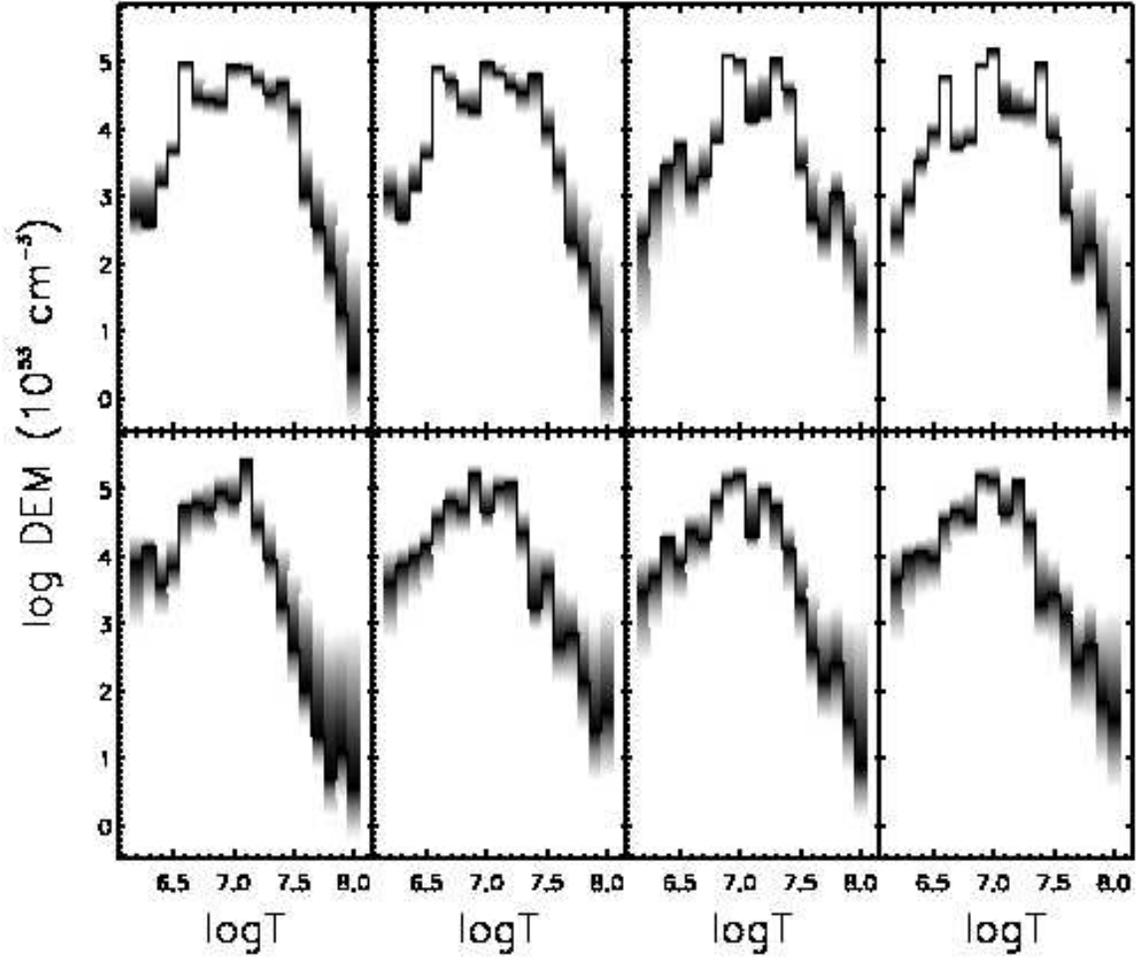}
%\end{turn}
\caption{Top panels: AB Dor DEM obtained by running a MCMC[M]
reconstruction code on a set of lines of H-like, He-like and highly
ionized Fe line fluxes. The panels from left to right correspond to
line set 1 (O, Ne, Mg, Si), set 2 (O, Ne, Mg, Si and S), set 3 (O, Si,
S, \ion{Fe}{17}, \ion{Fe}{18} and \ion{Fe}{21}) and set 4 (O, Ne, Mg,
Si, \ion{Fe}{17}, \ion{Fe}{18} and \ion{Fe}{21}). The thick solid line
represents the best-fit DEM, while the shaded regions correspond to
the 1-$\sigma$ deviations present in each temperature bin. Bottom
panels: Same as top panels but for V471 Tau.}
\label{f:dems}
\end{figure}

\begin{figure}[t]
\plotone{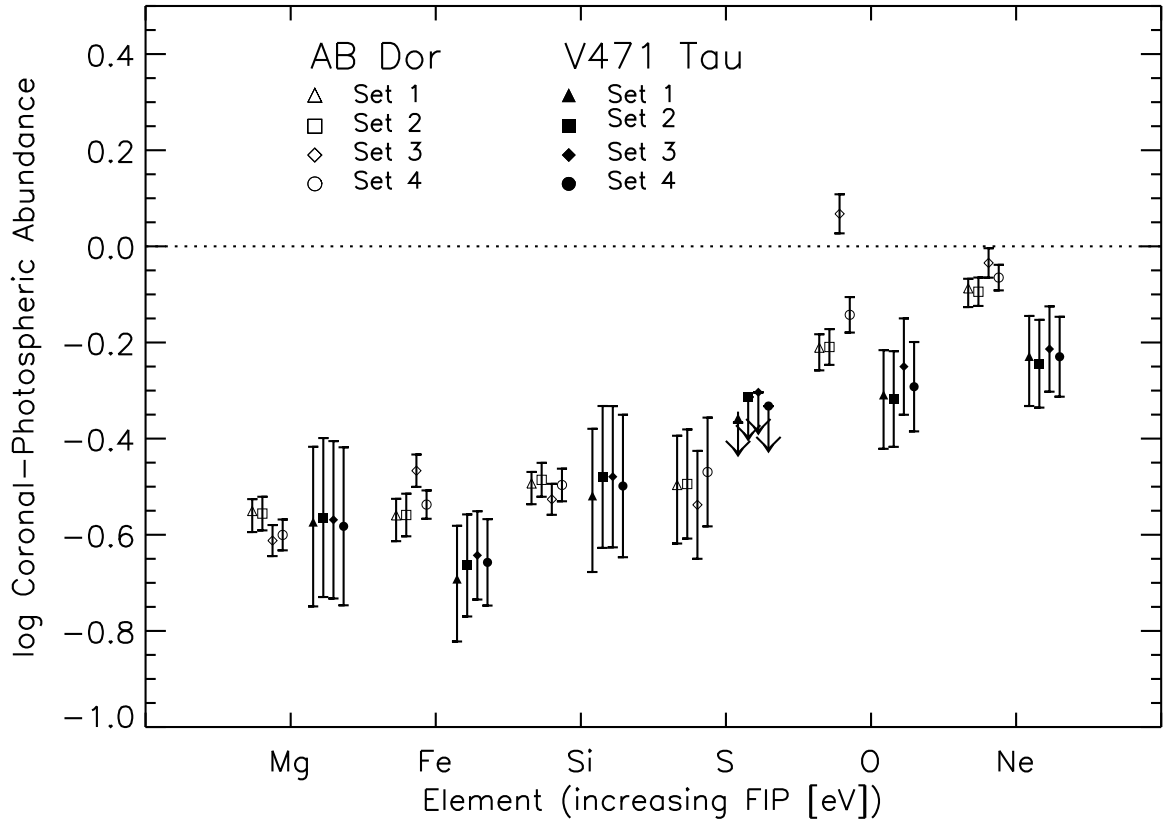}%abundance_FIP_singleplot_v2.ps}
\caption{\small{Coronal abundances obtained from abundance-insensitive
DEM reconstruction for AB Dor (open symbols) and V471 Tau (filled
symbols), relative to stellar photospheric values.  Note: The
abundances have been calculated using DEMs derived from four different
sets of ions for each object.}  The error bars represent statistical
uncertainties only: true uncertainties are likely to be approximately
0.1~dex.}
\vspace{-.6cm}
\label{f:abund}
\end{figure}

\begin{figure}[t]
\plotone{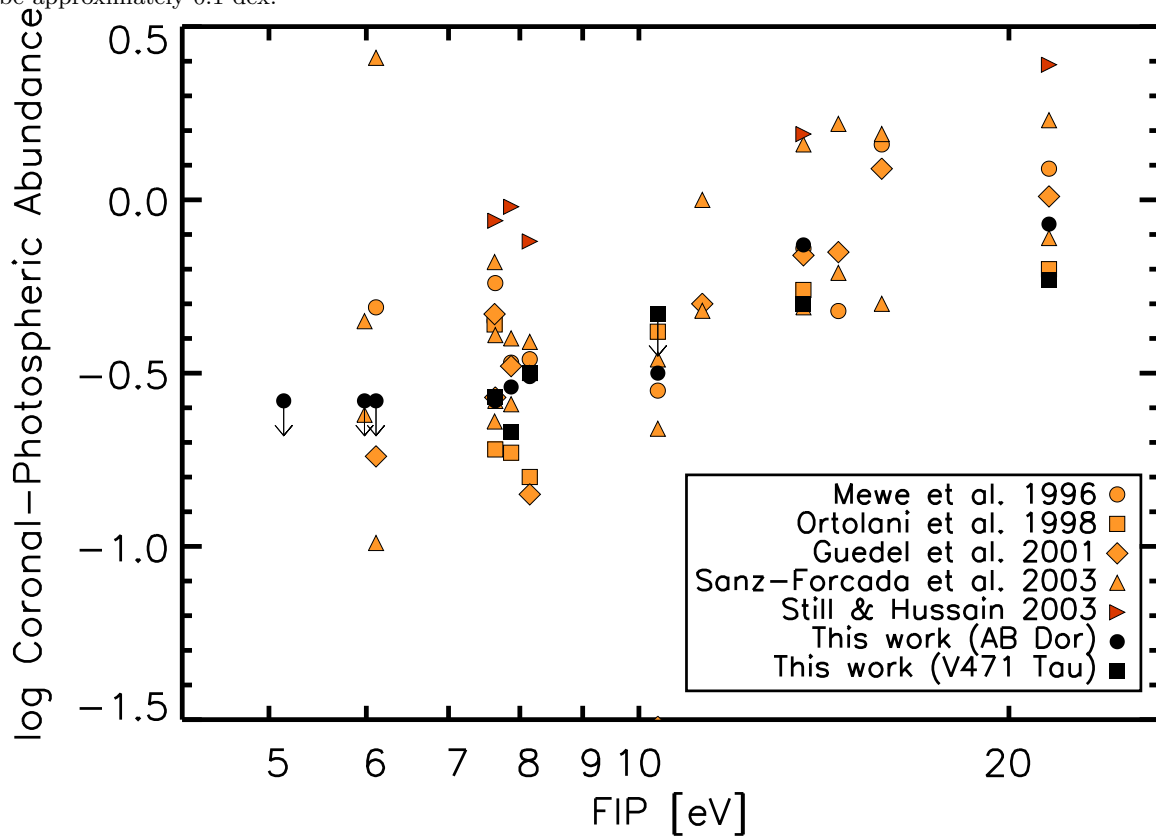}
\caption{\small{Comparison of coronal abundances against FIP from
AB~Dor and V471~Tau obtained from this work (black symbols) and the
literature (grey symbols). The uncertainties in the abundances derived
in this work are of order 0.1~dex. Note that the values from \citet{Guedel01} 
are from the RGS results.}}
\vspace{-.6cm}
\label{f:abund_liter}
\end{figure}

\begin{figure}[t]
\plotone{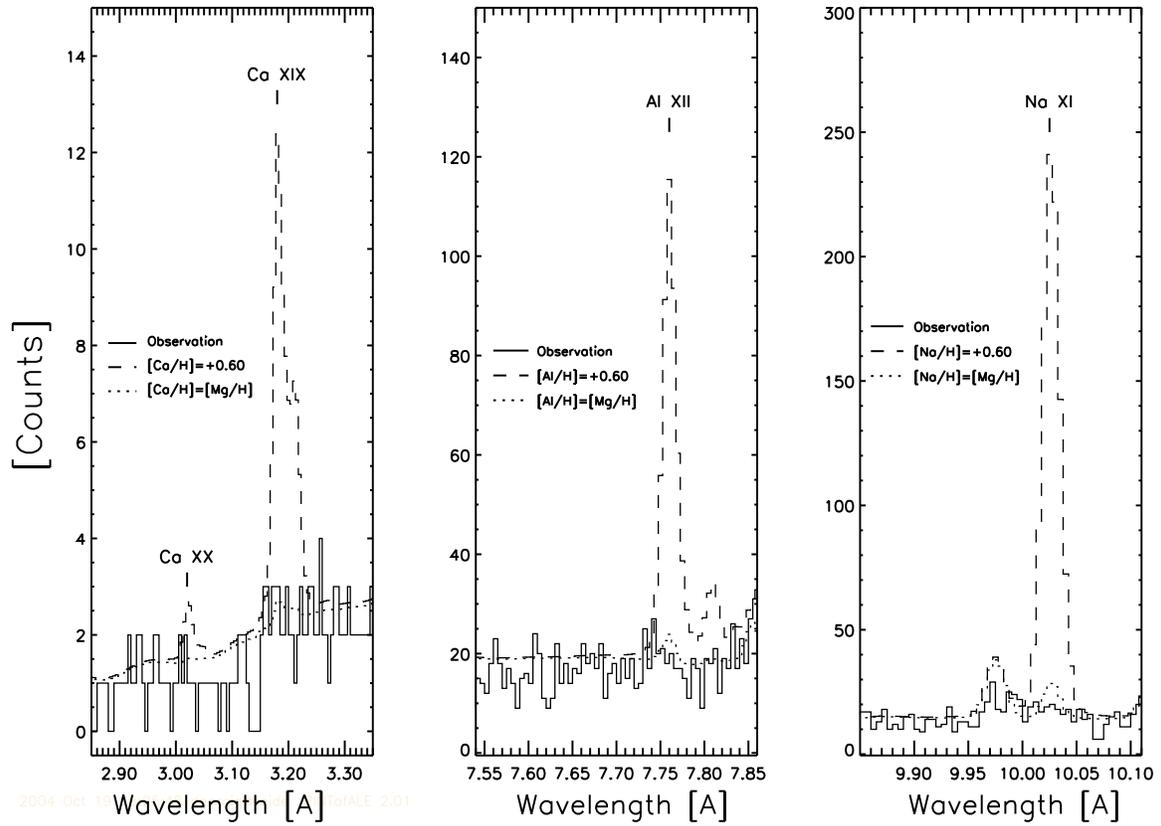}
\vspace{.2cm}
\caption{\small{Comparison between HETG+ACIS-S spectrum of AB Dor
(solid line) and the predicted spectrum derived for the H-like and
He-like Ca, Al and Na ions using two coronal abundances.}}
\label{f:low_fip_synt}
\end{figure}

\end{document}